\documentclass[aps,prx,superscriptaddress,twocolumn]{revtex4-1}

\usepackage{graphicx}
\usepackage{amsmath}
\usepackage{amssymb}
\usepackage[hidelinks]{hyperref}
\usepackage{color}
\usepackage{soul}
\usepackage{array}
\usepackage{braket}
\usepackage{bbold}
\DeclareMathOperator{\tr}{\text{Tr}}
\DeclareMathOperator{\diag}{diag}

\newcommand{\comments}[1]{}
\newcommand{\bea}{\begin{eqnarray}}
\newcommand{\eea}{\end{eqnarray}}

\newcommand{\ii}{\mathrm{i}}

\newcommand{\bs}{\boldsymbol}
\newcommand{\n}[1]{\,n\!\left(#1 \!\right)}

\usepackage{cancel}
\begin{document}

\title
{A quantum Otto engine with finite heat baths: energy, correlations, and degradation}

\begin{abstract}

We study a driven harmonic oscillator operating an Otto cycle by strongly interacting with two thermal baths of finite size. Using the tools of Gaussian quantum mechanics, we directly simulate the dynamics of the engine as a whole, without the need to make any approximations. This allows us to understand the non-equilibrium thermodynamics of the engine not only from the perspective of the working medium, but also as it is seen from the thermal baths' standpoint. For sufficiently large baths, our engine is capable of running a number of perfect cycles, delivering finite power while operating very close to maximal efficiency. Thereafter, having traversed the baths, the perturbations created by the interaction abruptly deteriorate the engine's performance. We additionally study the correlations generated in the system, and, in particular, we find a direct connection between the buildup of bath-bath correlations and the degradation of the engine's performance over the course of many cycles. 

\end{abstract}

\author{Alejandro Pozas-Kerstjens}
\affiliation{ICFO-Institut de Ciencies Fotoniques, The Barcelona Institute of Science and Technology, 08860 Castelldefels (Barcelona), Spain}

\author{Eric G. Brown}
\affiliation{ICFO-Institut de Ciencies Fotoniques, The Barcelona Institute of Science and Technology, 08860 Castelldefels (Barcelona), Spain}

\author{Karen V. Hovhannisyan}
\affiliation{Department of Physics and Astronomy, Ny Munkegade 120, Aarhus University, DK--8000 Aarhus, Denmark}
\affiliation{ICFO-Institut de Ciencies Fotoniques, The Barcelona Institute of Science and Technology, 08860 Castelldefels (Barcelona), Spain}

%\pacs{}

%05.30.-d: Quantum statistical mechanics
%03.67.Bg: Entanglement production and manipulation
%05.70.-a: Thermodynamics
%84.60.-h: Direct energy conversion and storage
%03.67.Mn: Entanglement measures, witnesses, and other characterizations

\maketitle

\section{Introduction}

The second law of thermodynamics prohibits extracting mechanical work from systems in thermal equilibrium. Therefore, in order to obtain work, one has to have access to systems out of thermal equilibrium. The theoretically simplest out-of-equilibrium system is one composed by two subsystems that are each at individual equilibrium and at different temperatures. This is the traditional setup for a heat engine: a working medium (WM) reciprocating between two thermal baths, pumps heat from the hotter bath (at temperature $T_h$) to the colder one (at temperature $T_c$) and outputs work as a result. The ideal engine converts the internal energy of the hot bath into work with an efficiency given by Carnot's formula, $\eta_C=1-T_c/T_h$. The idealizations needed for the machine to operate at such an efficiency are that (i) the baths interact with the working medium weakly \cite{ll5, Alicki_1979}, (ii) the cycle is a quasiequilibrium process and hence it takes infinite time to complete \cite{ll5, Sekimoto_2000, Shiraishi_2016}, and (iii) the baths are infinitely large \cite{ll5, Reeb_2014, Woods_2015, Tajima_2017, Richens_2017}. It has to be noted, however, that the size of the working medium itself is of no relevance -- it can be anything from a two-level quantum system \cite{Scovil_1959, Geva_1992, Rossnagel_2016} to a giant steam engine \cite{Callen}.

Strictly speaking, conditions (i) and (ii) can never be satisfied: any interaction has finite strength and any process that can be observed takes finite time. In the generic setup where the bath is a many-body system with short-range interactions and the WM couples to it locally, the breakdown of (ii) entails the failure of (iii) even if the bath is infinitely large \footnote{Whenever we refer to ``infinitely large'' systems, we mean finite systems that are so large that their finiteness cannot be observed within the largest timescale involved in the discussion.}. Indeed, in such systems, the Lieb-Robinson bounds \cite{Nachtergaele_2009, Nachtergaele_2010} imply that, roughly speaking, the correlations spread with finite velocity. This means that, in finite time, the WM can have access to only a finite region of the bath (see Ref. \cite{Masanes_2017}, where this idea was brought to use for the first time). However, it should be emphasized that the said finite region gets re-thermalized by the rest of the bath, so this scenario is not entirely equivalent to a finite bath.

Despite the significant attention that finite-time \cite{Reitlinger_1929, Curzon_1975, Velasco_1997, Sekimoto_2000, Van_den_Broeck_2005, Allahverdyan_2008, Seifert_2011, Correa_2013, Allahverdyan_2013, del_Campo_2014, Plastina_2014, Binder_2015, Shiraishi_2016, Benenti_2017}, strong-coupling \cite{Martinez_2013, Gallego_2014, Freitas_2014, Gelbwaser-Klimovsky_2015, Esposito_2015, Seifert_2016, Uzdin_2016, Strasberg_2016, Newman_2017, Perarnau-Llobet_2018, Benenti_2017}, and finite-size \cite{Allahverdyan_2011, Izumida_2014, Reeb_2014, Pekola_2016, Scharlau_2018, Tajima_2017, Masanes_2017, Richens_2017} effects have been getting either one by one or in groups of two, a rigorous microscopic analysis of a finite-power thermal machine strongly coupled to finite-sized heat baths has never been carried out. In this work, we aim to fill this gap by performing a fully microscopic analysis of a heat engine consisting of a harmonic oscillator serving as a WM, reciprocating---by being alternately strongly coupled and decoupled---between two finite, initially thermal harmonic chains serving as thermal baths.

The WM interacts with the baths via a modulated linear coupling (see Sec.~\ref{sec:interaction_with_single_bath} for details). This type of system-reservoir interaction is known under the name of Caldeira-Leggett model \cite{Caldeira_1983}, and is routinely used in many areas of physics ranging from quantum Brownian motion to quantum optics \cite{Weiss_1999, bp}. 

The engine runs a strong-coupling adaptation of the Otto cycle \cite{Callen}: the two ``isochoric'' thermalizations are intermediated by two ``adiabatic'' changes of the WM's Hamiltonian (see Sec.~\ref{sec:Otto} for the precise description). For the first cycle, the WM starts uncoupled from the baths and at equilibrium with the cold bath. This makes the initial state of the overall system a Gaussian state. Given that the total Hamiltonian is quadratic at any moment of time, the dynamics of the system can be described within the formalism of Gaussian quantum mechanics (GQM) \cite{Adesso_2007}. The latter maps the intractable Schr\"odinger equation in the infinite-dimensional Hilbert space of the overall system onto a linear evolution of the finite-dimensional phase space. This allows us to perform a comprehensive analysis of the machine's operation without the need to adhere to any of the many approximations usually made when dealing with quantum open-system dynamics \cite{Weiss_1999, bp}. Moreover, by directly simulating the overall system's evolution, we gain access to the states of the baths at any moment of time, which allows us to reveal the physical mechanisms governing the degradation and eventual exhaustion of the initial disequilibrium provided by the baths in the finite-size, finite-time, and strong-coupling regime. With our approach, we can easily work with baths of size up to 300 times the size of the WM with just a standard table-top computer.

The paper is organized as follows. First, in Sec.~\ref{sec:GQM}, we give a short account on the notions from GQM that will be needed throughout the rest of the paper. This section is intended as an introduction and can be safely omitted by those familiar with GQM. In Sec.~\ref{sec:interaction_with_single_bath}, we describe the interaction of the WM with a single bath. In Sec.~\ref{sec:Otto}, we explore the physics of the Otto cycle, focusing first on the performance of the cycle (Sec.~\ref{sec:Otto:performance}), and then on the dynamics and the role of correlations (Sec.~\ref{sec:Otto:correlations}). Finally, we summarize our conclusions in Sec.~\ref{sec:conclusions}. The codes, both in \textsc{Matlab} and Python, of all the numerical computations performed in this work are available in Ref.~\cite{compapp}.

\section{Review of Gaussian quantum mechanics} \label{sec:GQM}

In this section, we review the formalism of Gaussian quantum mechanics, focusing on the aspects necessary for our study. For a much broader introduction to the topic, the reader is referred to Ref.~\cite{Adesso_2007}. Note that throughout this paper all expressions are given in natural units, i.e., we assume $\hbar=k_B=1$.

The primary computational advantage of this formalism is that it allows us to study interacting systems via a direct system-plus-bath perspective, without having to resort to perturbation theory \cite{Brown_2013_nonperturbative} or other open-systems techniques. This provides access to the exact evolution of the bath in addition to the system, a fact we take great advantage of in this work.

Consider one or more quantum systems ascribed with bosonic canonical quadrature operators, satisfying the canonical commutation relations (CCRs), $[q_i,p_j] = \ii \, \delta_{ij}$, where the indices label the systems (henceforth referred to as oscillators or modes). If one were to think about a harmonic oscillator with Hamiltonian $\frac{P^2}{2\mu}+\frac{\mu\omega^2 Q^2}{2}$, then a convenient choice of quadratures would be $q=Q\sqrt{\omega \mu}$ and $p=\frac{P}{\sqrt{\omega \mu}}$. In terms of the creation and annihilation operators, the quadratures are expressed through $q_i = (a_i+a_i^\dagger)/\sqrt{2}$ and $p_i = \ii (a_i^\dagger - a_i)/\sqrt{2}$. For a system of $N$ modes, the quadratures form a phase space that we represent as the vector of operators
\begin{align}
    \bs x = (q_1,p_1, \cdots, q_N,p_N)^\text{T}.
\end{align}

Due to the CCRs, the phase space is a symplectic space, endowed with the structure \mbox{$[x_a,x_b]=\ii \, \Omega_{ab}$}. $\Omega_{ab}$ are the components of the so-called \textit{symplectic form}, given by
\begin{align} \label{symform}
	\bs \Omega = \bigoplus_{i=1}^N
	\begin{pmatrix}
		0 & 1 \\
		-1 & 0
	\end{pmatrix}.
\end{align}

In GQM one works with Gaussian states. A state of an $N$-mode system is Gaussian if and only if it is an exponent of a quadratic form in $\{x_a\}_{a=1}^{2N}$. Importantly, thermal states of quadratic Hamiltonians fall within this class. The defining feature of Gaussian states is that they are fully described by the first and second moments of their quadratures, i.e., their mean position and their variances in phase space. The mean quadratures of all the states we consider in this work will be zero, and so the formalism further simplifies. We thus characterize the state of our system via the $2N \times 2N$ covariance matrix $\bs \sigma$, the entries of which are given by
\begin{align} \label{covmat}
    \sigma_{ab} = \braket{x_a x_b + x_b x_a} = \tr\left[\rho\,(x_a x_b + x_b x_a)\right].
\end{align}

An important aspect of GQM is that creating ensembles and performing partial traces is trivial. This is due to working in phase space rather than in a Hilbert space, where partitions are represented as a direct sum rather than as a tensor product. Thus, any combined state of two systems $A$ and $B$ takes the form
\begin{align}  \label{twomodestate}
	\bs{\sigma}_{AB}=
	\begin{pmatrix}
		\bs{\sigma}_A & \bs{\gamma}_{A B} \\
		\bs{\gamma}_{A B}^\text{T} & \bs{\sigma}_B
	\end{pmatrix},
\end{align}
where $\bs \sigma_A$ and $\bs \sigma_B$ are the reduced states of systems $A$ and $B$ respectively, and the matrix $\bs \gamma_{AB}$ specifies the correlations between the systems. The superscript $\text{T}$ denotes the operation of transposition.

A fact crucial for GQM is that any unitary evolution generated by a time-dependent Hamiltonian that is quadratic at any moment of time will preserve the Gaussianity of a state \cite{Schumaker_1986}. Any such unitary, $U$, on the Hilbert space corresponds to a linear symplectic transformation on the phase space of quadratures: $\bs x \rightarrow U^\dagger \bs{x} U=\bs{S}\bs{x}$, with $\bs{S}$ satisfying
\begin{align} \label{symplectic}
	\bs{S} \bs{\Omega} \bs{S}^\text{T}=\bs{S}^\text{T} \bs{\Omega} \bs{S}=\bs{\Omega}.
\end{align}

The symplecticity of $\bs{S}$, expressed by Eq.~\eqref{symplectic}, ensures that the CCRs are preserved throughout the change of basis. On the level of the covariance matrix, it is easy to see that this transformation acts as
\begin{align}
	\bs{\sigma} \rightarrow \bs{\sigma'}=\bs S \bs \sigma \bs S^\text{T}.
\end{align}

\subsection{Energy, Evolution, and Thermality}

Another convenient aspect of GQM is that it allows us to compute average energies, evolve the system over time according to some time-dependent quadratic Hamiltonian, and diagonalize the system into its normal mode basis without ever referencing a Hilbert space object.

The average energy of a state represented by the covariance matrix $\bs \sigma$, with respect to a purely quadratic Hamiltonian $H = \bs x^\text{T} \bs F \bs x$, is given by
\begin{equation}     \label{energyEqn}
    \braket{H} = \frac{1}{2}\tr (\bs F \bs \sigma).
\end{equation}

The symplectic (i.e., unitary in the Hilbert space) evolution matrix $\bs S(t)$ generated by this (in general, time-dependent) Hamiltonian obeys a Schr\"odinger-like equation:
\begin{equation}  \label{evolutioneqn}
    \frac{d \bs S(t)}{dt} = \bs \Omega \bs F_s(t) \bs S(t),
\end{equation}
where $\bs F_s = \bs F+\bs F^\text{T}$. For a constant Hamiltonian the solution trivially takes the form $S(t)=\exp(\bs \Omega \bs F_s t)$, and for general driven systems, the equation can be straightforwardly integrated by standard numerical techniques.

When speaking of a ``free'' system we mean that we are working in the basis that diagonalizes the system's Hamiltonian. This is called the normal mode basis, in which the Hamiltonian takes the form
\begin{align} \label{Hfree}
    H_\text{free}=\sum_{i=1}^N \omega_i a_i^\dagger a_i = \sum_{i=1}^N \frac{\omega_i}{2}(p^2_i+q^2_i),
\end{align}
where, in the second equality, we have ignored the (constant) zero-point energy. The corresponding phase-space matrix is diagonal in this basis: \mbox{$\bs F_\text{free} =\tfrac{1}{2} \diag(\omega_1,\omega_1, \omega_2, \omega_2, \cdots)$}. By definition, the normal modes do not interact with each other. This means that any thermal state on the entire system is given by the tensor product (in phase space, the direct sum) of the individual normal modes' thermal states.

In general, the system may have couplings between pairs of modes (for example, between nearest neighbours), which give non-diagonal elements to the matrix $\bs F$. The normal-mode basis can be obtained by symplectically diagonalizing this matrix: $\bs S \bs F \bs S^\text{T} = \bs F_\text{free}$, where $\bs S$ is a symplectic matrix, and $\bs F_\text{free}$ is diagonal as above.

In the normal-mode basis the covariance matrices of the system's thermal states are given by
\bea \label{GroundThermal}
\bs \sigma_T = \bigoplus_{i=1}^N \begin{pmatrix}
	\nu^{(th)}_i & 0 \\
	0 & \nu_i^{(th)}
\end{pmatrix}, \quad \nu_i^{(th)} \!= \frac{e^{\omega_i /T}+1}{e^{\omega_i/T}-1},~~
\eea
where $\omega_i$ are the normal frequencies. We can thus find the thermal covariance matrix of any interacting system by first identifying the normal basis, specifying the covariance matrix $\bs \sigma$ as above, and then applying the inverse transformation to this matrix to put it back into the physical-mode basis.

The values $\nu_i^{(th)}$ in Eq.~\eqref{GroundThermal} are referred to as the thermal state's symplectic eigenvalues. In general, every Gaussian state of $N$ modes has $N$ symplectic eigenvalues $\nu_i$, which are obtained by symplectically diagonalizing the covariance matrix: there always exists a symplectic matrix $\bs S$ such that
\begin{align}
    \bs S \bs \sigma \bs S^\text{T} = \bigoplus_{i=1}^N
    \begin{pmatrix}
		\nu_i & 0 \\
		0 & \nu_i
	\end{pmatrix}.
\end{align}
The symplectic eigenvalues can be directly computed by taking the regular eigenvalues of the matrix $\ii \bs \Omega \bs \sigma$, which come in $\pm \nu_i$ pairs. 

\subsection{Entropy and Correlations}

Consider a two-party state of the form of Eq.~\eqref{twomodestate}. The off-diagonal matrix $\bs \gamma_{AB}$ contains the correlation functions between the two systems, and these systems are uncorrelated if and only if $\bs \gamma_{AB}=0$. As a measure of correlations we use the mutual information, defined as
\begin{align}
    I(A,B)=S(\bs{\sigma}_{A})+S(\bs{\sigma}_B) - S(\bs{\sigma}_{AB}).
\end{align}

Here, $S(\bs \sigma)$ is the von Neumann entropy of the state with covariance matrix $\bs \sigma$, given by
\bea \label{vnent0}
S(\bs{\sigma})=\sum_{i=1}^N f(\nu_i),
\eea
where
\bea \label{vnent1}
f(\nu) = \frac{\nu+1}{2}\log\frac{\nu+1}{2}-\frac{\nu-1}{2}\log\frac{\nu-1}{2}.
\eea
This shows that the symplectic eigenvalues of a state -- which are invariant under symplectic transformations -- give a measure of mixedness for that state. For example, the entropy is zero, i.e., a Gaussian state is pure, if and only if all its symplectic eigenvalues are equal to one. Note that no state can have eigenvalues smaller than one (this is a statement of the uncertainty principle).

We are thus able to very easily compute the mutual information across any partition in our system, independent of how many modes each partition contains.

Note that the entanglement is also computable, but, for most situations, it is considerably more difficult. In the particular case of two modes it is nevertheless easy \cite{Brown_2013}, and we discuss some findings in that regard in the next sections. However, due to the thermality of our system, quantum correlations are hard to maintain, and we have found that generally entanglement does not play a significant role in the scenarios we consider below. Interestingly, this aspect is in accord with (yet by no means logically necessitated by) the fact that, although capable of manifesting many interesting quantum features, GQM is an essentially classical, noncontextual sector of quantum mechanics in that it can be described by a local hidden variable model \cite{Bartlett_2012}.

\section{Gaussian interaction with a single bath} \label{sec:interaction_with_single_bath}

Before performing the analysis of the Otto cycle, let us study some relevant features of the isochoric interaction of the WM with a single thermal bath. We will thereby introduce the specific Hamiltonians that describe the components of the Otto engine in the next sections.

Throughout this work, we model thermal baths as collections of harmonic oscillators arranged in one-dimensional, translation-invariant rings with nearest-neighbour interactions. We consider only position-position couplings so that the free Hamiltonian of a bath is given by
\begin{equation}
    H_\text{bath}=\sum_{i=1}^{N}\frac{\omega_b}{2}\left(p_{i}^2+q_{i}^2\right) + \sum_{i=1}^N\alpha q_{i}q_{i+1},
    \label{Hbath}
\end{equation}
where $N$ is the number of oscillators in the bath, $\omega_b$ is the bare frequency of each of them, and $\alpha$ controls the coupling strength. Note that, because of the periodic boundary conditions, $q_{N+1}=q_1$.

The phase-space matrix corresponding to this Hamiltonian is
\begin{align}
    \bs F_\text{bath} = \frac{1}{2} \begin{pmatrix}
		\bs \omega_b & \bs \alpha & \bs 0 & \cdots & \bs 0 & \bs \alpha \\
		\bs \alpha & \bs \omega_b & \bs \alpha & \cdots & \bs 0 & \bs 0 \\
		\bs 0 & \bs \alpha & \bs \omega_b & \cdots & \bs 0 & \bs 0 \\
		\vdots & \vdots & \vdots & \ddots & \vdots & \vdots \\
		%\bs 0 & \bs 0 & \bs 0 & \cdots & \bs \omega_h & \bs \alpha \\
		\bs \alpha & \bs 0 & \bs 0 & \cdots & \bs \alpha & \bs \omega_b
	\end{pmatrix},
\end{align}
where $\bs 0$ is the $2\times 2$ matrix of zeros, and
\begin{align}
    \bs \omega_b = \begin{pmatrix}
        \omega_b & 0 \\
        0 & \omega_b
        \end{pmatrix}, \;\;\;
    \bs \alpha = \begin{pmatrix}
        \alpha & 0 \\
        0 & 0
        \end{pmatrix}.
\end{align}

At the beginning of the process, the bath is initialized in a thermal state $\rho(0)\propto e^{-H_\text{bath}/T_b}$ at temperature $T_b$. 

Due to the interactions, the covariance matrix will \emph{not} be given by a simple direct sum as in Eq.~\eqref{GroundThermal}. Rather, we must first identify the normal mode basis that symplectically diagonalizes the Hamiltonian matrix $\bs \omega_\text{bath}/2 = \bs S \bs F_\text{bath} \bs S^\text{T}$, where $\bs \omega_\text{bath}$ is the diagonal matrix composed of the normal mode frequencies of $H_\text{bath}$. We then identify the thermal state as per Eq.~\eqref{GroundThermal} and finally transform back to the physical basis to find the thermal state of the ring, $\bs \sigma_\text{bath} = \bs S^{-1} \bs \sigma_T (\bs S^{-1})^\text{T}$ (this calculation is included in the function \textit{Initialize} in Ref.~\cite{compapp}).

As a WM we employ yet another harmonic oscillator, with bare frequency $\omega_m$. 
Its coupling to the bath is described by
\begin{equation}
    H_{int}=\gamma \lambda(t) \, q_m \sum_{i \in \{\text{int}\}} q_i 
    % = \gamma \lambda(t) q_m \, q_1,
    \label{Hint}
\end{equation}
where $\{\text{int}\}$ is the set of the bath's nodes which the WM interacts with.

For most of our analysis, we choose this set to contain just one and always the same node of the bath, which we label the \textit{first node} $q_1$. However, also a situation where the interaction set has more than one element is discussed in Sec~\ref{sec:Otto:performance}.

The function $\lambda(t)$ is a \textit{switching function} that modulates the interaction in time. In particular, we choose the following compactly-supported, smooth switching function
\begin{equation}
    \lambda(t)=\begin{cases}
    0 & t < 0 \\
    \frac{1}{2}-\frac{1}{2}\tanh\cot\frac{\pi t}{\delta} & 0\leq t < \delta\\
    1 & \delta \leq t < \tau-\delta \\
    \frac{1}{2} + \frac{1}{2}\tanh\cot\frac{\pi (t-\tau)}{\delta} & \tau-\delta \leq t <\tau \\
    0 & t > \tau
    \end{cases},
    \label{swaping}
\end{equation}
where $\tau\geq 2\delta$ is the total duration of the interaction with the bath and $\delta$ is the time that takes to fully switch on (and fully switch off) the interaction. We will refer to $\delta$ as the \textit{ramp-up time} of the isochoric interaction.

The phase-space matrix of the overall Hamiltonian will then be $\bs F_\text{tot}=\left(\begin{array}{cc}
\tfrac{1}{2}\bs \omega_m & \tfrac{1}{2}\bs{\gamma} \\
\tfrac{1}{2}\bs{\gamma}^\text{T} & \bs F_\text{bath}
\end{array}\right)$. Here $\bs \gamma$ is a $2\times 2N$ matrix containing all zeros except for the first entry $\gamma_{11}=\lambda(t)\gamma$, corresponding to the $q_m q_1$ interaction we are imposing.

In order to construct the symplectic evolution matrix $\bs S(t)$ of the overall system, which is generated by $\bs F_\text{tot}$, we numerically integrate Eq.~\eqref{evolutioneqn}. The covariance matrix at the moment $t$ will then be simply given by \mbox{$\bs \sigma_\text{tot}(t) = \bs S(t) (\bs \sigma_m \oplus \bs \sigma_\text{bath}) \bs S(t)^\text{T}$}, where \mbox{$\bs \sigma_m = \text{diag}(\nu^{(m)},\nu^{(m)})$} is the initial state of the WM (its values being given by Eq.~\eqref{GroundThermal}) \cite{compapp}.

\begin{figure}
    \centering
    \includegraphics[width=0.48\textwidth]{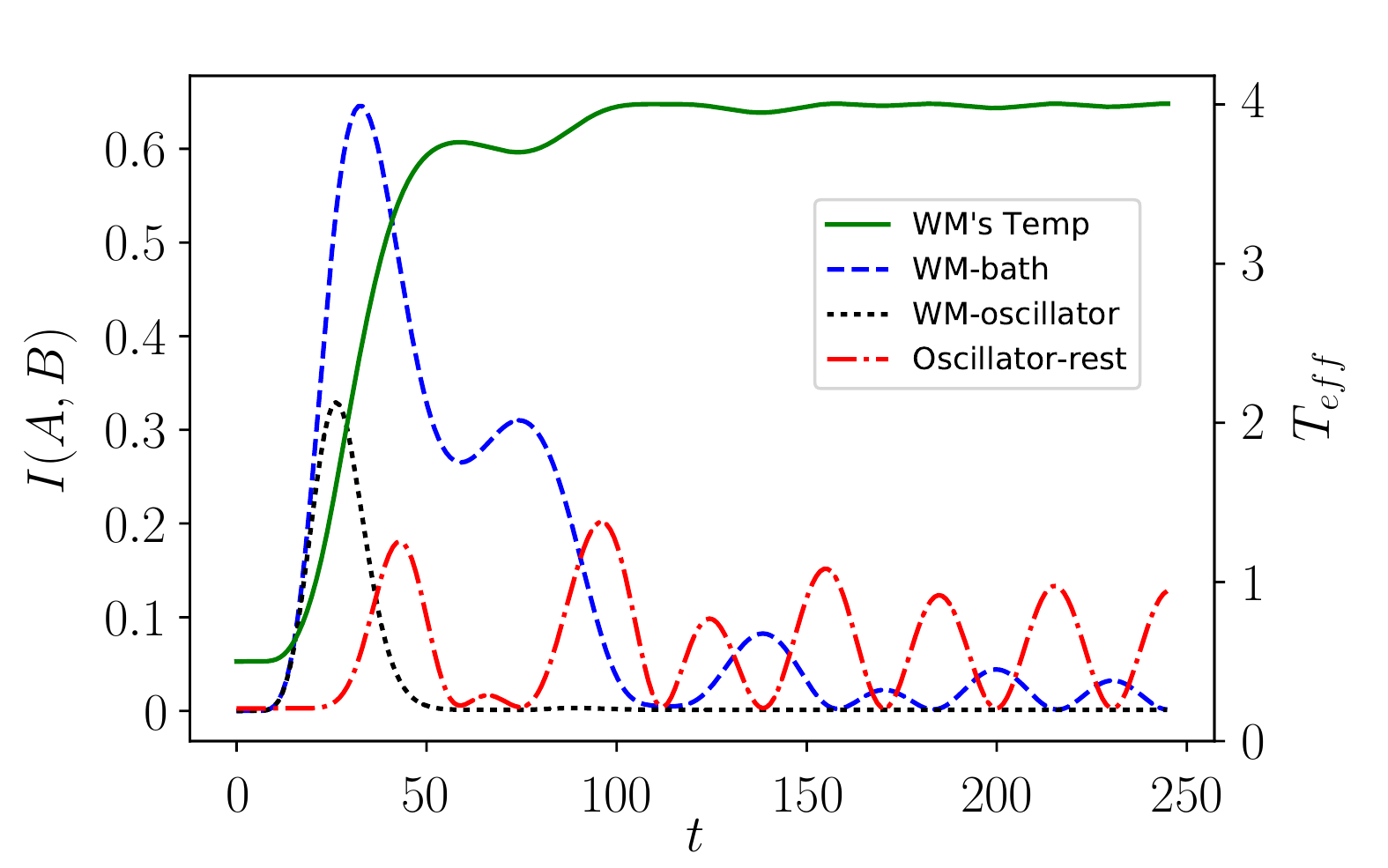}
    \caption{Evolution of several quantities during a period of interaction between the WM and a thermal bath. The green solid line is the WM's effective temperature. The other lines represent the mutual information between various partitions: (dashed blue) between the WM and the bath as a whole, (black dotted) between the WM and the specific bath oscillator with which it interacts, and (red dot-dashed) between this oscillator and the rest of the bath. The bath contains $N=30$ oscillators, all with frequency $\omega_b=2$, and is initialized in a thermal state at temperature $T_b=4$. The WM has also frequency $\omega_m=\omega_b$, but is initialized in a thermal state at temperature $T_m=0.5$. The total time of interaction is $\tau=245$, the ramp-up time is $\delta=0.1\tau$, and the interaction strengths are $\alpha=\gamma=0.1$. For these parameters, the exact thermalization time as defined in the main text is $\tau_{th}=98.5$. Note that the red curve is not initially zero (and it should not be, because there is initial correlation from the ring couplings). However, since we are working with a relatively hot bath, these correlations are very small (of the order of $10^{-3}$), and its magnitude cannot be appreciated in full detail in the figure.}
    \label{fig:transfer}
\end{figure}

During the evolution of the overall (WM plus bath) system, we have found that the state of the WM remains very close to being thermal. That is, at any moment of time, its covariance matrix is very close to that given by Eq.~\eqref{GroundThermal} for some $\nu^{(th)}$ (for a detailed discussion and an explicit characterization of the distance of the actual state of the WM to a thermal state, see Appendix~\ref{app:temperature}). Given this, we are able to assign a meaningful effective temperature to the WM by computing the temperature associated with its symplectic eigenvalue. An example of this is shown by the green, solid line of Fig.~\ref{fig:transfer}.

Importantly, we notice that at $t \approx 93$ the temperature of the WM becomes equal to that of the bath. We note that this moment is not the thermalization time in the proper sense because the interaction is still on. Rather, the exact thermalization time is $\tau_{th}\approx 98.5$. It turns out that, in order to achieve precise thermalization, one needs to match the frequencies, so that $\omega_m$ is also the frequency of the individual bath oscillators (the $\omega_b$ in Eq.~\eqref{Hbath}), and the couplings, so that the WM-baths interaction strength is equal to the intra-ring coupling strength, i.e., $\gamma=\alpha$. Intuitively, this matching ensures that the rate of information transfer between the WM and the bath is the same as between oscillators within the bath. Whenever $\omega_m$ (resp. $\gamma$) is outside a small neighbourhood of $\omega_b$ (resp. $\alpha$), the WM does not thermalize with the bath at all (similar frequency filtering phenomenon in the classical setting was reported in \cite{Smith_2008}). In view of this, we from now on set $\omega_m=\omega_b$ and $\gamma=\alpha$.

With such a configuration of the parameters, and fixed $\delta/\tau$, the thermalization time $\tau_{th}$ scales as
\bea \label{relaxtime}
\tau_{th}\propto \alpha^{-1}
\eea
for $\alpha\ll 1$. In fact, the above scaling is, to a good approximation, preserved also for large $\alpha$. A remark is in order here. As $\tau_{th}$, we choose the smallest $\tau$ that achieves thermalization. Since the bath is finite and the WM couples to it strongly, the temperature of the WM will not approach the bath's temperature $T_b$ in a monotonic way: with passing time, the WM's final temperature will first go slightly above $T_b$ -- the maximum being $T_b + \mathcal{O}(\alpha^2)$ \footnote{This is due to the fact that, because of the interaction, the local effective temperature of a bath node is slightly above the global temperature} -- then go below $T_b$, and continue an oscillatory behavior as that depicted in Fig.~\ref{fig:transfer} for $t>100$.

Another important aspect of our thermalization process is that, due to the finite duration and the finite strength of the WM-bath interaction, it has a non-zero work cost. More specifically, the extracted work, as quantified by the difference between the initial and final average energies of the total system, is not zero. However, despite the strong non-equilibrium character of the process, this amount is small (compared to, e.g., the energy exchanged between the WM and the bath). In fact, for small $\alpha$, this work cost, $W_i$ (where the subscript $_i$ stands for isochoric), scales as
\bea
W_i\propto \alpha^2,
\eea
and is almost independent of the ramp-up time, $\delta$. Taking, for example, $N=30$, $\gamma = \alpha = 0.1$, $\omega_m = \omega_b = 2$, $T_b = 4$, $T_m = 0.5$, $\tau = 100$, and $\delta=0.1\tau$, we get $W_i \approx -6.2\times 10^{-3}$, while the exchanged heat is $\approx 3$. This, together with the fact of exact thermalization discussed above, means that the fine-tuning of the frequencies and couplings provides us with an example of almost work-free thermalization in finite time, resulting from a strong interaction between the WM and the bath. A similar example, where the structure of the bath is known and the Hamiltonian of the WM is finely-tuned, was constructed in \cite{Allahverdyan_2010}. This is not a standard, exponential relaxation behaviour \cite{bp}, and it can be argued that such behaviour cannot occur for general baths of unknown structure \cite{Allahverdyan_2013}. 

We also note that the fact of almost zero work justifies the usage of the term ``isochoric'' for this process. Indeed, in this case, most of the energy exchange is heat transfer, which is the characteristic of isochoric processes \cite{Callen}. In the strong coupling regime, strictly isochoric (or, equivalently, constant-Hamiltonian processes) cannot exist as any non-zero coupling will change the system Hamiltonian, and therefore the term needs to be adapted.

Furthermore, subtle processes such as the evolution of the correlations between the WM and the bath or information exchange between the WM and the bath can be examined in very great detail within the framework of GQM. As an illustration, in Fig.~\ref{fig:transfer} we examine the evolution of the correlations between various partitions during the interaction.

We compare the correlations, as measured by the mutual information, between the WM and the whole bath (dashed blue line), between the WM and the node in the bath it interacts with (black dotted line), and between the latter and the rest of the bath (red dot-dashed line). This provides us with a number of insights into the non-perturbative interaction of the WM and the bath. First, during the phase of switching on the interaction, the WM and the bath quickly build up strong correlations, which decay later on. This decay is caused by the fact that the bath nodes to which the WM is coupled also interact with the rest of the bath, and the bath, due to its tendency to thermalize, forces these correlations to decay. We can see this process in more detail by examining the other two lines. The correlation between the machine and the interacting node similarly rises and then falls, and the decay occurs exactly as this node becomes significantly correlated with the rest of the bath.

This gives us an important intuitive picture. The interaction between the WM and the bath generates correlations between the two (specifically, between the WM and the interacting node). Due to the intra-bath couplings, the WM also becomes correlated with other ring nodes in an outwards-propagating manner. However, these couplings also mean that the correlation between the WM and bath will, over time, be swapped to correlations between different bath nodes, as we see, for example, in the red dot-dashed line in Fig.~\ref{fig:transfer}. Over the course of many interaction sessions, the bath nodes therefore become more and more intercorrelated, which will eventually result in a halt of the machine. We elaborate on this process in the next section, where we discuss the performance of a WM operating cyclically between two finite-sized baths.

\begin{figure}
    \centering
    \includegraphics[width=0.48\textwidth]{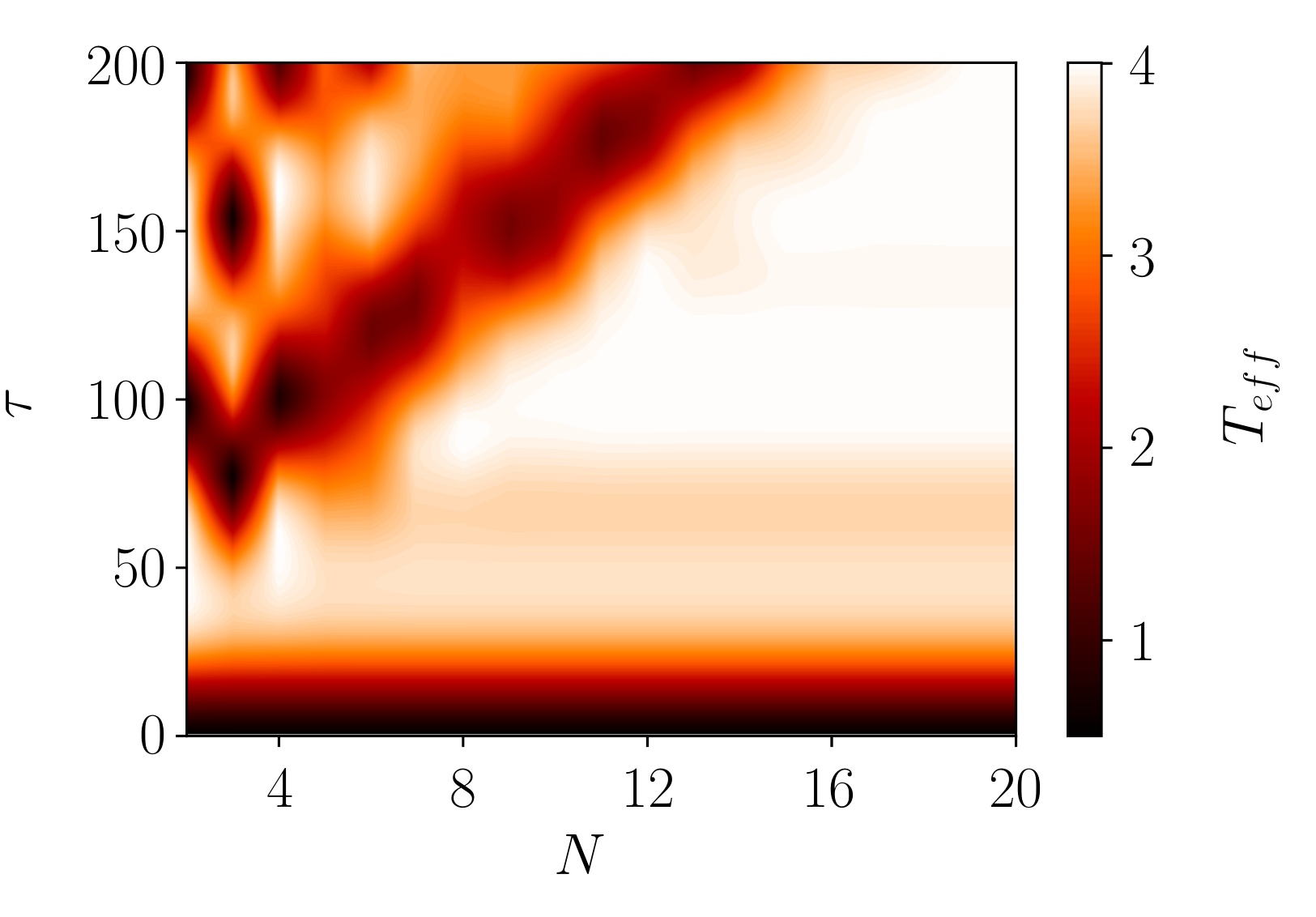}
    \caption{Effective temperature of the WM after the interaction with the bath as a function of the bath size $N$ and the time of interaction $\tau$. The parameters for the WM are those used also for Fig.~\ref{fig:transfer} ($\omega_m=2$, $T_m=0.5$), and similarly for the relevant parameters of the bath ($\omega_b=2$, $T_b=4$, $\alpha=\gamma=0.1$). The ramp-up time of the interaction is $\delta=0.1\tau$ for every value of $\tau$. Note the two distinct behaviors separated by a straight line $\tau=c\cdot N$.}
    \label{fig:lightcone}
\end{figure}

One does not need to move to the two-bath scenario to observe the effects of having finite-sized baths, though. In fact, one only needs to interact with the bath for a time that is long enough. We do so in Fig.~\ref{fig:lightcone}, where we compute the effective temperature of the machine after interacting with a bath composed of $N$ nodes during a time $\tau$, for different values of $N$ and $\tau$, and all the other parameters being the same as those used for Fig.~\ref{fig:transfer}.

In Fig.~\ref{fig:lightcone}, we observe two very distinct behaviors that are clearly separated. For $\tau<c\cdot N$, where $c$ indicates the slope of the ``causal cone'', the temperature of the WM is insensitive to the size of the bath. Indeed, the interaction time in this case is short enough so as to allow the interaction to finish before the perturbations that propagate through the bath return to the region which interacts with the WM (i.e., the interacting node of the ring). Therefore, there is no difference between the temperature that the WM achieves in this case and the temperature that it would achieve from interacting with an infinite bath. The opposite occurs for $\tau>c\cdot N$: in this case, the interaction time is long enough so as to permit the perturbations generated by the interaction with the WM to return to the interacting node. These perturbations modify the local state of the interacting node, which in turn translates into a response in the WM that diverges from that expected for infinite baths.

It is also worth noting that, for short interaction times, the WM does not have enough time to fully thermalize with the bath. We observe that the effective temperature of the WM increases with the interaction time until the point where thermalization is achieved. After this point, increasing the interaction time further has no major influence on the WM's temperature until, of course, it is long enough for the perturbations to go around the bath. 

\section{The Gaussian Otto cycle}
\label{sec:Otto}

\begin{figure}
    \centering
    \includegraphics[width=0.49\textwidth]{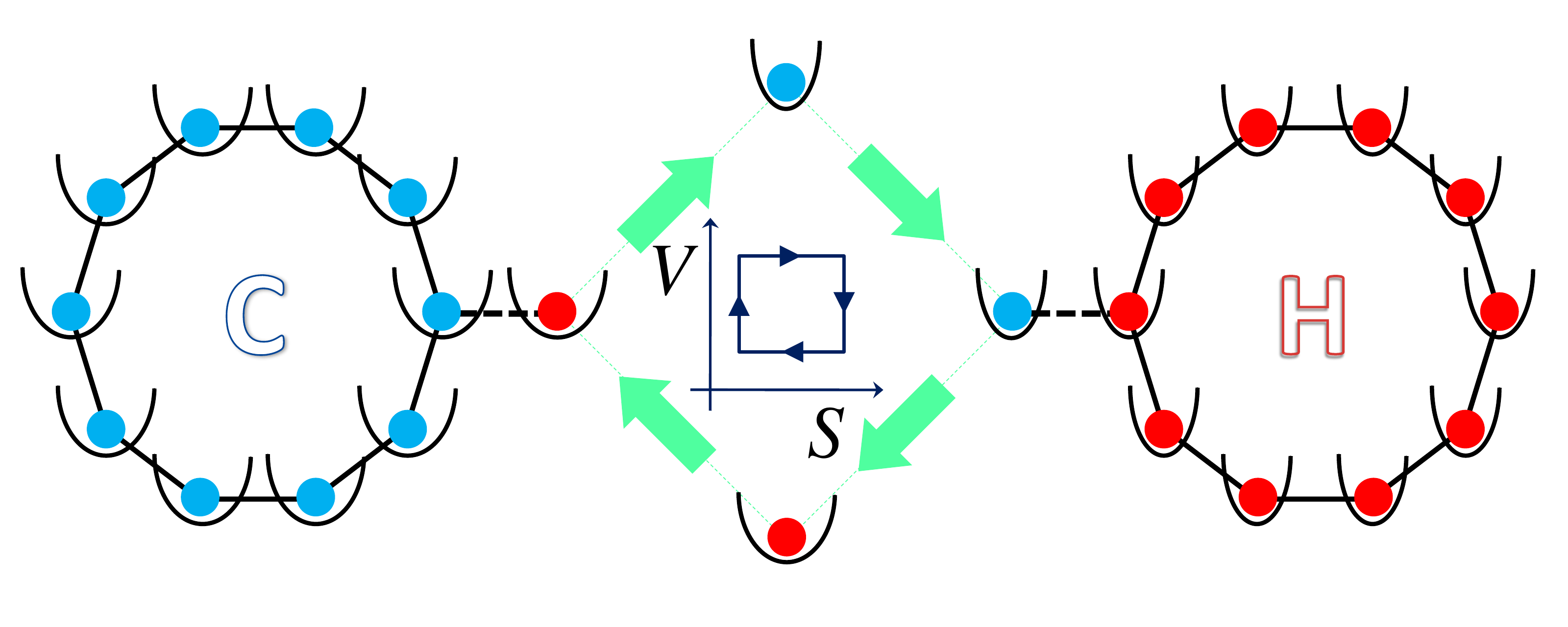}
    \caption{Visualization of the Otto cycle. The standard sequence of isochoric thermalisations and adiabatic compressions/expansions defining the Otto cycle in phenomenological thermodynamics are, in our case, implemented as a sequence of $q-q$ interactions (as described in Sec.~\ref{sec:interaction_with_single_bath}) and sudden changes of the WM's Hamiltonian. More specifically, the cycle consists of the following steps: (i) the WM interacts with the hot bath (the red harmonic chain) by a coupling that is smoothly switched on, kept constant, and smoothly switched off, (ii) the Hamiltonian of the WM is suddenly changed so that the frequency matches the individual frequencies in the cold bath, (iii) the WM is brought into contact with the cold bath (the blue harmonic chain), with the same pattern of interaction as in step (i), and (iv) the Hamiltonian of the WM is suddenly changed back to its original value.}
    \label{fig:Otto}
\end{figure}

We now study the performance of the WM running an Otto cycle between two thermal baths at temperatures $T_h$ (hot) and $T_c$ (cold), as depicted in Fig.~\ref{fig:Otto}.

\begin{figure*}[t!]
    \centering
    \begin{tabular}{ccc}
    \includegraphics[width=0.46\textwidth]{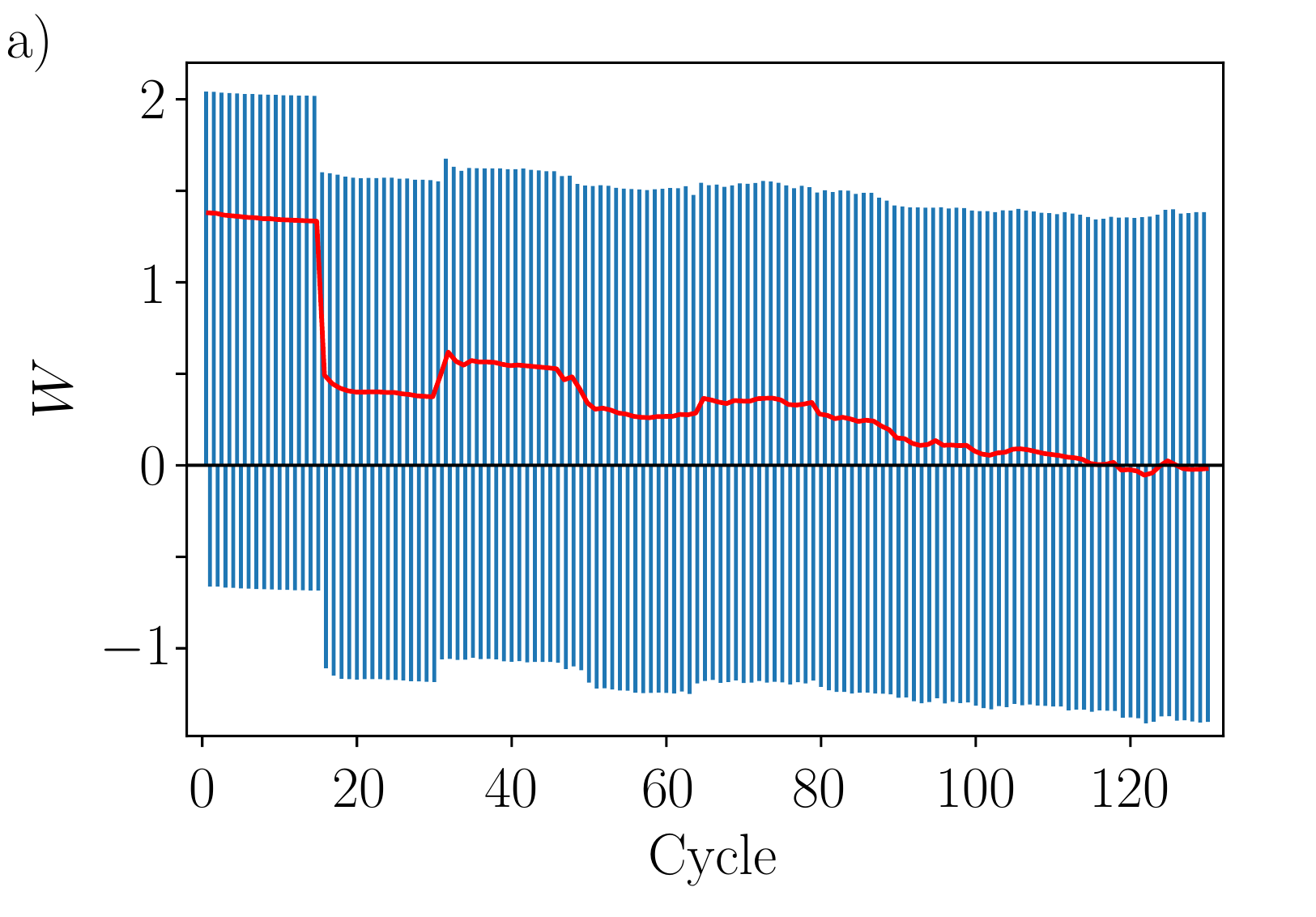} &
    \includegraphics[width=0.5\textwidth]{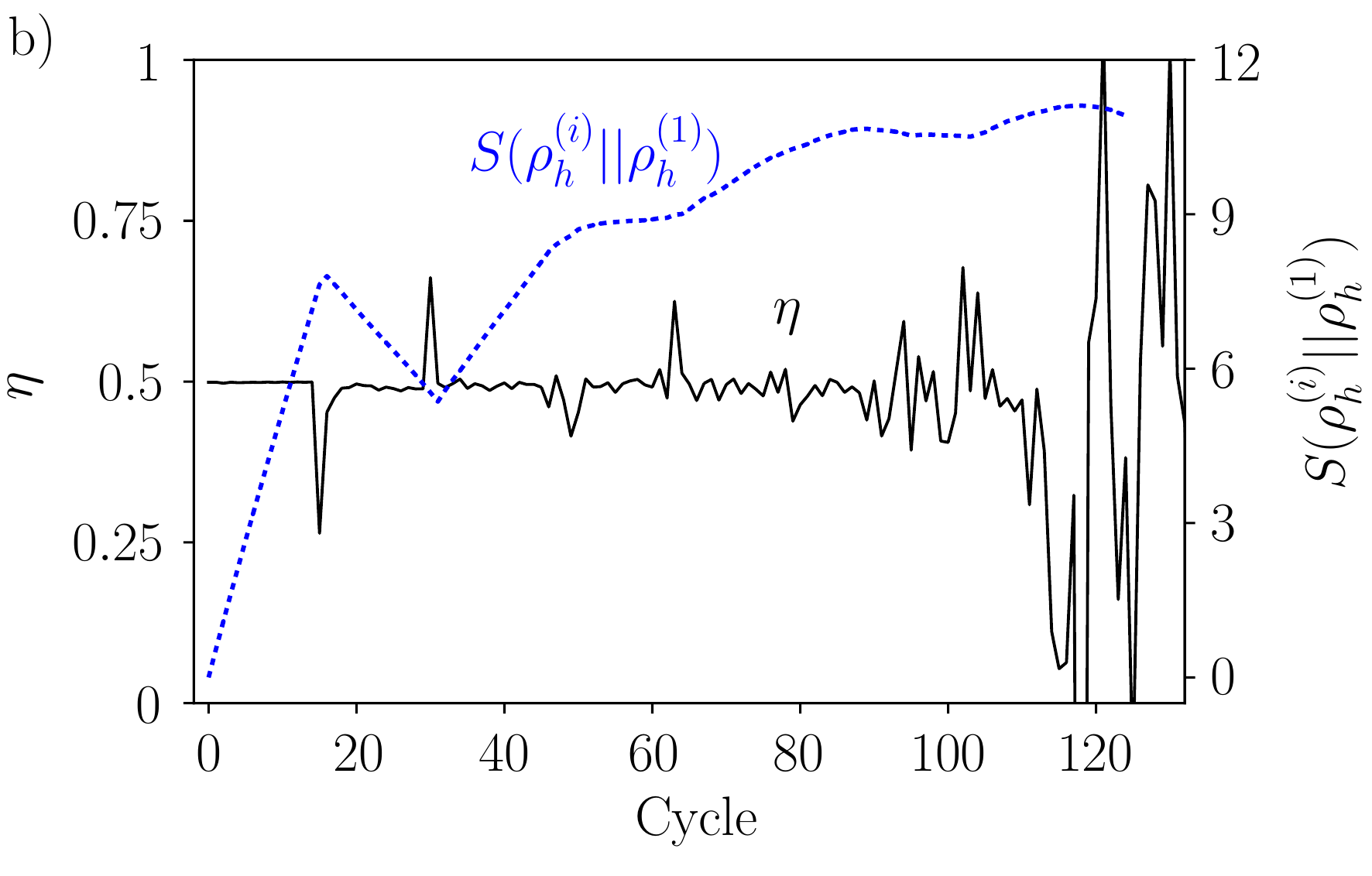}
    \end{tabular}
    \caption{a) Work output during the adiabats. One pair of positive-negative bars represents a full cycle. The red, solid line represents the total work extracted from each cycle (positive bar + negative bar + work during the isochores). The relevant parameters of the system are $N_h=N_c=300$, $T_h=4$, $T_c=0.5$, $\omega_h=2$, $\omega_c=1$, $\alpha_h=\alpha_c=\gamma=0.1$, $\tau=100$, $\delta=0.1\tau$. b) (Solid black) Efficiency of the engine and (dotted blue) relative entropy of the hot bath with respect to its initial state for every cycle of operation. The divergences in the efficiency are caused by the extraction of no heat from the hot bath.  Note how the efficiency and work output of the engine remaining virtually constant during the ``perfect'' cycles is contrasted by the steady increase of the relative entropy distance of the bath's state from its initial value.}
    \label{fig:wadiabatic}
\end{figure*}

The Otto cycle we consider is composed of two isochoric interactions between the machine and each of the baths (as described in Sec.~\ref{sec:interaction_with_single_bath}) separated by two sudden changes of the WM's Hamiltonian. Specifically, between subsequent interactions, we instantaneously swap the WM's Hamiltonian,
\begin{equation} \label{swap}
    H_m=\omega_c a^\dagger_m a_m \leftrightarrow H_m' = \omega_h a_m^\dagger a_m,
\end{equation}
so that the WM's state remains unchanged. The fact that the WM is detached from the baths during the swap ensures that the process is adiabatic, i.e., thermally isolated, in the thermodynamic sense \footnote{Moreover, the fact that the eigenbasis of the Hamiltonian remains unchanged additionally ensures that the process is adiabatic also in the sense of the quantum adiabatic theorem.}. It is important to note that the change in Eq.~\eqref{swap} is not equivalent to simply quenching the frequency of the oscillator. Rather, it requires simultaneously changing both the mass and the frequency: $\mu\to \mu\frac{\omega_c}{\omega_h}$ and $\omega_c\to\omega_h$. Here, $\omega_c$ and $\omega_h$ are the frequencies of the WM used during the interactions with the cold and hot baths, respectively. For a discussion of the case when only the frequency is quenched, see Appendix~\ref{app:swap}. As mentioned in Sec.~\ref{sec:interaction_with_single_bath}, $\omega_c$ and $\omega_h$ are chosen to coincide with the frequencies of the nodes of, respectively, the cold and the hot baths. Moreover, we also match the interaction strength with the ring coupling strengths, i.e., $\gamma=\alpha_c=\alpha_h$ (we choose $\alpha_c=\alpha_h$ for simplicity only, without losing generality).

The total work extracted during a cycle is given by the sum of works extracted during each of the four parts of the cycle. As we showed in Sec.~\ref{sec:interaction_with_single_bath}, the work contributions from WM-bath interactions are small, hence most of the work is generated during the adiabats. The work produced by a sudden change in Hamiltonian in Eq.~\eqref{swap} is given by a particularly simple expression. Indeed, since the baths remain intact during the adiabat, the work is given by the energy change of the WM only. For example, in the adiabat after an interaction with the hot bath, the energy of the WM is decreased by \mbox{$W_{h\to c} = (\omega_h - \omega_c)\tr \bs \sigma_m$}. If we choose $\omega_c<\omega_h$, not only will $W_{h\to c}>0$, but also the net work will be positive. Note also that if $\omega_c>\omega_h$ we would be running a refrigerator.

Lastly, note that the engine cycles are not cyclic in the standard thermodynamic sense. Indeed, since the baths are finite and the interaction with WM perturbs them non-negligibly, at the end of each cycle, the state of the WM (and, of course, the state of the baths) will be different from that at the beginning. Nevertheless, the deviation from cyclicity is small during the period of ``perfect'' cycles that we describe below.

\subsection{Cycle performance}\label{sec:Otto:performance}

We begin the cycle by the interaction with the hot bath, so the starting Hamiltonian of the WM is $\omega_h a_m^\dagger a_m$ and its state is thermal, at temperature $T_c$ with respect to $\omega_c a_m^\dagger a_m$. Due to the finite size of the baths and the strong perturbations that the interactions with the WM causes in them, we expect that the performance of the engine will drop over time. This intuition is confirmed in Fig.~\ref{fig:wadiabatic}, where we plot the work output and efficiency of the engine as a function of the number of cycles of operation. In Fig.~\ref{fig:wadiabatic}a, each bar represents the work output during an adiabat, and the red line represents the total work output in each cycle (the sum of the works in the adiabats plus the sum of the works in the isochores), as described above. The heat $Q$ is defined as the energy the hot bath loses per cycle. We define the energy of the bath with respect to the Hamiltonian in Eq.~\eqref{Hbath}, and, for the $n$-th cycle, the heat is given by
\bea
    Q = \!-\Delta E_h = \tr \left[\bs F_\text{bath} (\bs \sigma_h(2n\tau)\!-\!\bs \sigma_h((2n\!+\!1)\tau) )\right],~~~~
\eea
where $\bs \sigma_h (t)$ is the covariance matrix describing the state of the hot bath as a function of time. The efficiency of the engine is defined as usual: $\eta=W/Q$.

In Fig.~\ref{fig:wadiabatic}, we see that the engine's performance has two regimes. First, the work output and absorbed heat are approximately constant, decreasing very slowly, for the first 15 complete cycles. We call these ``perfect'' cycles. The degradation of the engine's performance during these cycles is due to the residual perturbations near the interaction site that the outward-propagating perturbations created by the WM-bath interaction leave behind. As the cycles proceed, these small deviations from the interaction site's equilibrium state accumulate, causing the gradual decrease in work and heat.

What is more, during the perfect cycles, the perturbations, created by the WM-bath interaction, propagate through the baths in the same way as they would do were the baths infinite. Therefore, any given perfect cycle is unaffected by the further increase in the size of the baths. This implies that the engine's degradation in the course of perfect operation is not a finite-size effect, and hence occurs also when the baths are infinite. Moreover, the difference between the work outputs in, say, the first and second cycles, does not vanish when the coupling is taken to zero. Therefore, the perfect-regime degradation is not a strong-coupling effect either. Rather, it strongly depends on the ramp-up time and can be decreased noticeably by increasing $\delta/\tau$. However, going to high values of $\delta/\tau$ prevents the WM from thermalizing with the baths, thereby impairing the functioning of the engine. Hence, the degradation cannot be eliminated completely in our model. We explicitly compute the correction to the optimal figures of merit due to this degradation in Appendix~\ref{app:eff}. We note that, while the dependence of single-cycle characteristics on the ramp-up time is in line with the general intuition that non-commutative, time-dependent interactions generate excitations that cause thermodynamic friction (see, e.g., \cite{Rezek_2006, Plastina_2014}), the important fact of the cycle-to-cycle accumulation of the imperfections caused by finite switching time is a separate phenomenon.

We furthermore observe that the number of perfect cycles, $N_p$, increases asymptotically linearly with \mbox{$N\equiv N_c=N_h$}, the number of nodes in the baths, as is to be expected given the constant, finite speed of propagation of the perturbations in the bath \footnote{Recall that, for nearest-neighbour Hamiltonians such as that in Eq.~\eqref{Hbath}, the Lieb-Robinson bounds set limitations on the propagation of perturbations \cite{Nachtergaele_2009, Nachtergaele_2010}}. However, when the interaction time $\tau$ is close to $\tau_{th}$, $N_p$ does not depend on $\alpha$ for small $\alpha$. Indeed, although the thermalization time increases with decreasing $\alpha$ (see Eq.~\eqref{relaxtime}) and this requires longer interaction times $\tau$ with the baths, the propagation of perturbations within the baths also slows down, and the two effects almost exactly compensate each other. Along with the fact that the degradation is slow, the linear dependence of $N_p$ on $N$ makes the perfect regime relevant for practical engines with large baths.

Differences from the perfect-cycle behavior begin to appear only when the perturbations return to the region of the bath that directly interacts with the WM. This is the point at which the finite-size effects take relevance, and it is marked by the drastic, discontinuous drop in the work output in Fig.~\ref{fig:wadiabatic}. The performance of the engine becomes unreliable due to large variations that heat and work undergo both in magnitude and sign. The above discontinuous behaviour of the engine's figures of merit is contrasted with the conventional gradual degradation of the performance of an engine operating between finite reservoirs (see, e.g., \cite{Izumida_2014}). The contrast is further sharpened by the observation that, as is also the case in the said conventional picture, the baths diverge from their initial states in a gradual, continuous manner. This is illustrated in Fig.~\ref{fig:wadiabatic}b, where the distance, as measured by relative entropy \cite{mikeike}, of the hot bath's state at the beginning of the $i$-th cycle, $\rho_h^{(i)}$, to the bath's initial state, $\rho_h^{(1)}$, $S(\rho_h^{(i)}||\rho_h^{(1)})=\tr\left[\rho_h^{(i)}\left(\ln \rho_h^{(i)} - \ln \rho_h^{(1)}\right)\right]$ (see Appendix~\ref{app:relentropy} for a more detailed discussion), is plotted as a function of $i$. In Fig.~\ref{fig:wadiabatic}b it can be seen how, during the ``perfect'' cycles, the characteristics of the engine stay almost constant despite the fact that the bath's state changes at a steady rate.

An important implication of Fig.~\ref{fig:wadiabatic} is that, during the perfect cycles, the efficiency of the engine $\eta$ is very close to $\eta_O = 1 - \omega_c/\omega_h$. The latter is the theoretical maximum for an oscillator running an idealized Otto cycle between two infinite thermal baths to which it is coupled weakly enough for the standard Markovian open quantum system techniques \cite{bp} to be applicable \cite{Rezek_2006, Quan_2007, Kosloff_2017}. Such idealized engines are known to obey the so-called power-efficiency trade-off, which states that the power output of the engine has to approach to zero whenever the efficiency comes close to the reversible maximum (see, e.g., \cite{Shiraishi_2016, Kosloff_2017, Allahverdyan_2013}). Our model respects the power-efficiency trade-off for the Otto cycle in the following manner: for $\alpha\ll1$, the efficiency approaches $\eta_O$ from below as
\bea \label{eficiencia}
\eta_O -\eta \propto \alpha^2,
\eea
while for the work output of a perfect cycle we have \mbox{$W=W_{\alpha=0} - \mathcal{O}(\alpha^2)$}. We refer the reader to Appendix~\ref{app:eff} for a more detailed discussion on these quantities. Taking into account Eq.~\eqref{relaxtime}, this leads us to
\bea \label{poder}
P\propto \alpha.
\eea
Here $P$ is the power output of the engine: \mbox{$P=W/\tau_\text{cycle}$}, where $\tau_\text{cycle}=2\tau$ is the duration of the cycle. We note that, although the setting of our problem is different from that in Ref.~\cite{Perarnau-Llobet_2018}, the scalings in Eqs.~\eqref{eficiencia} and \eqref{poder} agree with (and saturate) the optimal scalings derived there.

One can also consider coupling the WM to more than one, evenly spaced ring sites. It turns out that adding more interacting sites reduces the amount of perfect cycles, which matches the intuitive picture described earlier. Indeed, the reduced distance between the sites leads to shorter time needed for the perturbations created by the interaction to reach the nearest site of interaction. Interestingly, the work output of a single perfect cycle is insensitive to the cardinality of the set $\{\text{int}\}$ (as long as the perturbations generated in one interacting site do not have time to arrive to any other), but of course the total work output of the engine over several cycles does get reduced by increasing the number of interaction points. On the other hand, the more sites the WM interacts with, the smaller is the time necessary for it to thermalize. This leads to an increased power output for the initial perfect cycles, albeit at the cost of decreasing the number of such cycles.

\subsection{Propagation of correlations}
\label{sec:Otto:correlations}

As noted before, the formalism presented in Sec.~\ref{sec:GQM} allows for an easy way of identifying whether two systems are correlated. In this subsection, we use this property to study how correlations distribute along the baths and the WM. We consider this as one of the (probably many) paths to obtain a better understanding of the phenomenology presented above.

\begin{figure}[h]
    \centering
    \includegraphics[width=0.48\textwidth]{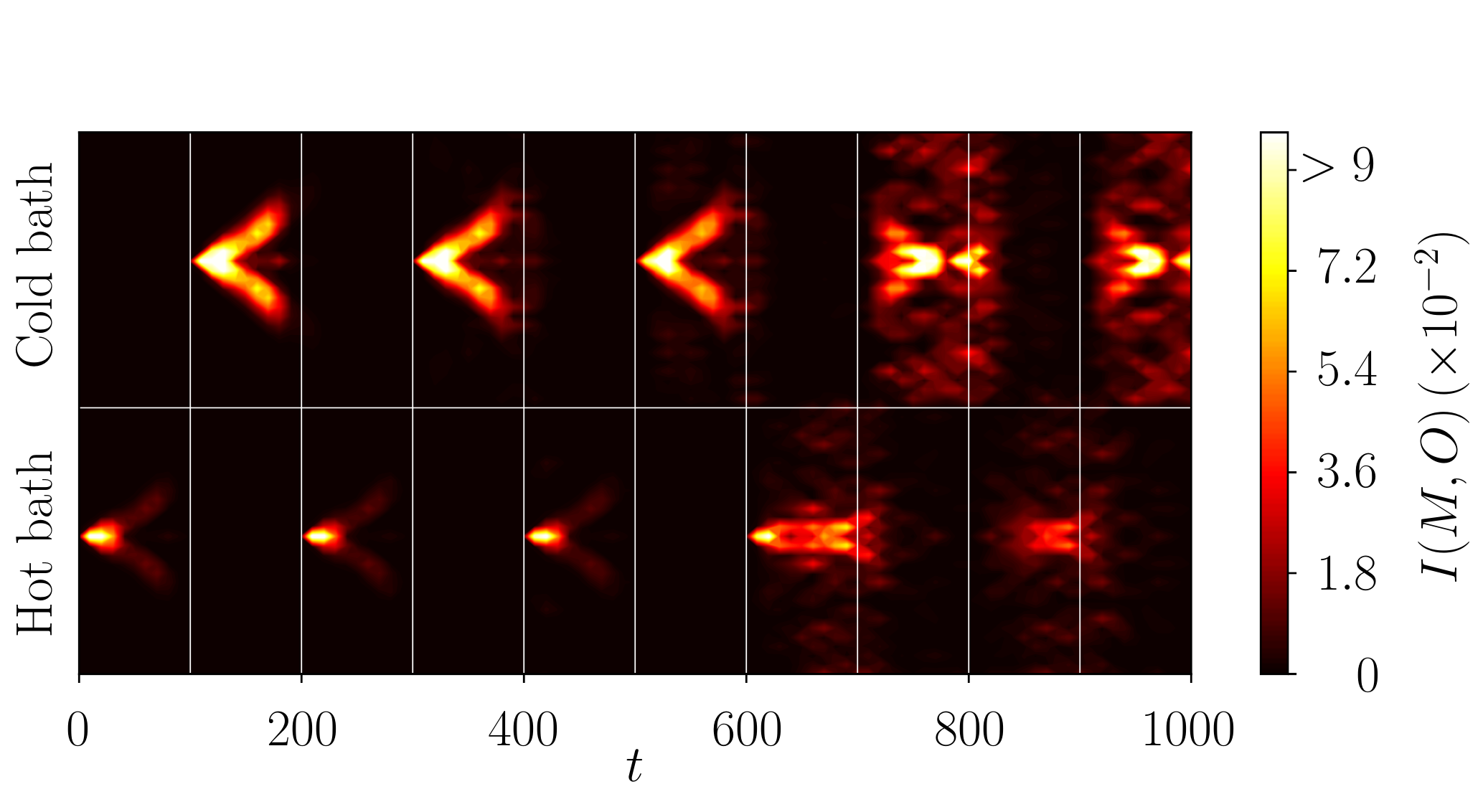}
    \caption{Mutual information between the WM and each oscillator in each bath during the five initial cycles of operation. The horizontal line separates the two baths, and the vertical lines separate the interactions of the machine with each of the baths. The baths have $N=30$ oscillators each, and are initially uncorrelated and at temperatures $T_c=0.5$ and $T_h=4$. The rest of relevant parameters are as those used in Fig.~\ref{fig:Otto}. Note how during the first three cycles the machine observes no differences from the interaction with infinite baths, and after this point the perturbations in the chains arrive back to the interacting oscillator, modifying its local state.}
    \label{fig:detbathcorrelations}
\end{figure}

In Fig.~\ref{fig:detbathcorrelations}, we show the strength of the correlations between the WM and each of the oscillators in each bath and how these correlations evolve in time for five consecutive cycles. Throughout this subsection, each bath is composed of $N=30$ oscillators, with all other parameters being the same as in Sec.~\ref{sec:Otto:performance}. The vertical lines denote the instants of time at which the machine stops interacting with one bath and, after the corresponding adiabat, begins interacting with the other. 

One feature we immediately observe is the explicit propagation of the perturbations in the form of localized wavepackets at finite speed, in full agreement with the Lieb-Robinson bound \cite{Nachtergaele_2009,Nachtergaele_2010}. Although it is hard to see in Fig.~\ref{fig:detbathcorrelations}, while propagating, these wavepackets leave residual perturbations behind. The latter are small and, during the first three cycles of operation ($t\in [0,\,600]$), the WM appears to interact with almost unperturbed baths. These are the perfect cycles described above. The time $t=600$ is when the perturbations that were generated during the first three cycles manage to intercept the WM as it is \emph{currently interacting} with the bath, leading to the sudden drop of the work output that has been discussed in Sec.~\ref{sec:Otto:performance}.

We also observe that the propagating correlations quickly fade. This is unsurprising, and carries the same explanation as that given for Fig.~\ref{fig:transfer}. Our computations show that, to a surprisingly good approximation, during an interaction, the WM becomes correlated with just a single non-local mode in the bath---the mode that propagates outwards---as can be appreciated in Fig.~\ref{fig:detbathcorrelations}. However, both the WM and this propagating mode are interacting with the rest of the bath as well, and thus this correlation is quickly lost and distributed among bath modes. This also explains why the decay occurs much faster in the hot bath than in the cold bath. Indeed, the hotter the bath, the larger the thermal noise that will break the correlations.

\begin{figure}[h]
    \includegraphics[width=0.45\textwidth]{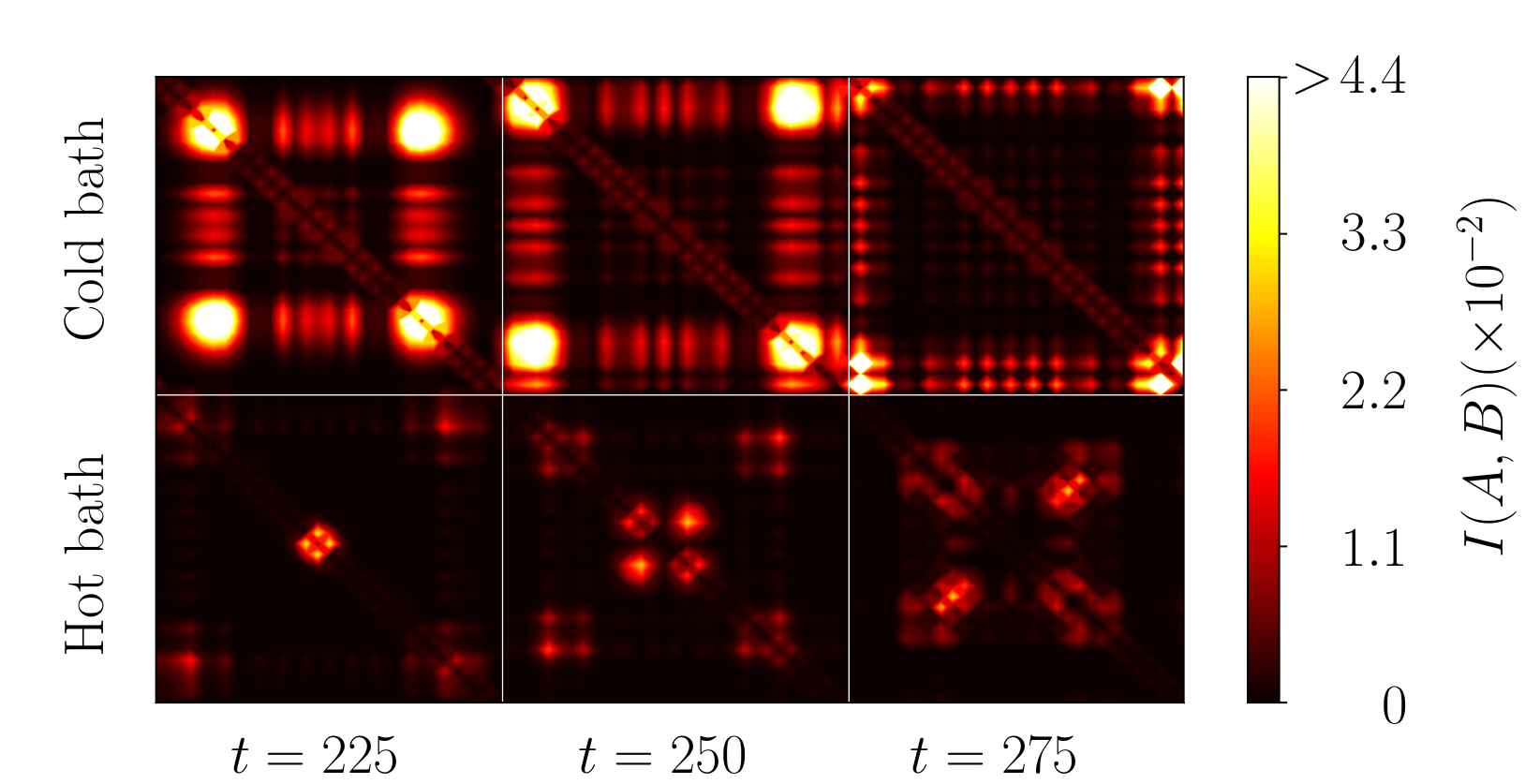}
    \caption{Intra-bath correlations at different moments of time, during the second cycle of operation of the engine. The parameters chosen are the same as those in Fig.~\ref{fig:detbathcorrelations}. Note that in the cold bath the correlations propagate outwards, away from the interaction point (in the center of the images), while in the hot bath we observe two waves: one propagating outwards ---generated by the latest interaction with the WM--- and one propagating inwards, generated in the previous interaction of the bath and the WM. The latter wave, due to the boundary conditions of the system, returns to the interaction point. The script \textit{correlations} in the computational appendix~\cite{compapp} generates a full animation of this phenomenon.}
    \label{fig:bathbathcorrelations}
\end{figure}

It is also instructive to observe how the correlations are built up and distributed along the baths. This is illustrated in Fig.~\ref{fig:bathbathcorrelations}, in which the mutual information between each pair of oscillators in each bath is shown at various times. One immediately notices the outward-propagating nature of these correlations.

\begin{figure}[h]
    \includegraphics[width=0.48\textwidth]{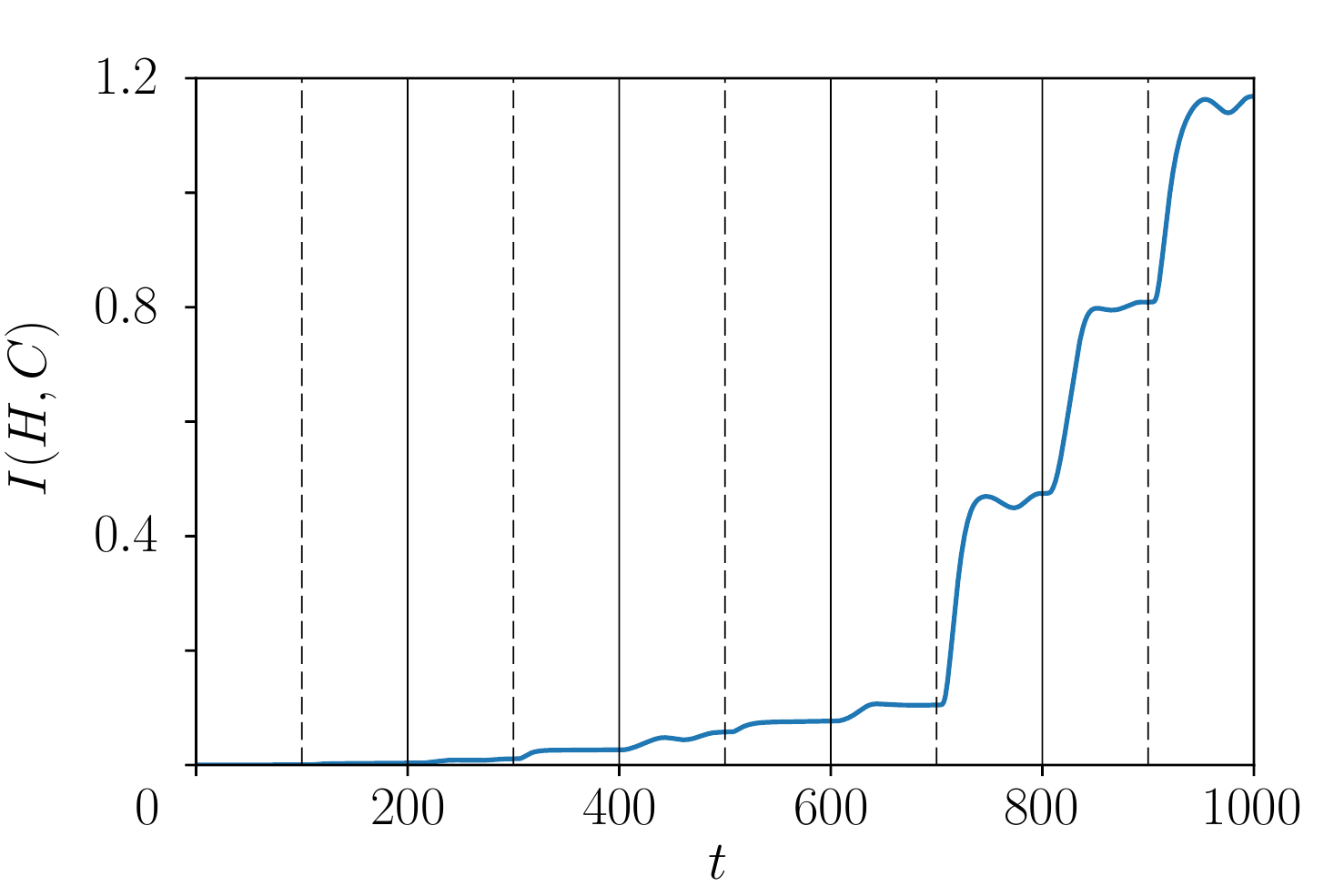}
    \caption{Mutual information between the hot and cold baths during the five initial cycles of interaction. The parameters used are the same as for Figs.~\ref{fig:detbathcorrelations} and \ref{fig:bathbathcorrelations}. The solid lines denote the end of each cycle, while the dashed lines denote the end of the interaction of the WM with the hot bath and the beginning of its interaction with the cold bath. Note the abrupt increase of the mutual information during the fourth cycle (starting at $t=600$), which is the first cycle outside the regime of ``perfect'' operation.}
    \label{fig:bathsmutualinfo}
\end{figure}

An important insight into the process of the engine's degradation is gained by looking at the bath-bath correlations instead. Indeed, given that the WM acts as a carrier of both energy and correlations between the baths, and that the baths gradually evolve away from their initial states, one would expect that, over time, the baths get more and more correlated and end up reaching a global passive state (see Ref.~\cite{Brown_2016} for the characterization of passivity within GQM). We explore this intuition in Fig.~\ref{fig:bathsmutualinfo}, in which we show that, surprisingly, the mutual information between the two baths remains close to zero during the ideal cycles, and starts abruptly increasing after the last ideal cycle is complete. This can be explained by noticing in Fig.~\ref{fig:detbathcorrelations} that, during the perfect cycles, the WM is virtually uncorrelated with the baths both at the beginning and at the end (but not in the middle) of each interaction session, which means that the WM does not transmit correlations during these cycles. This picture obviously breaks down once the perturbations reach the interaction site. This thereby establishes a clear quantitative link between the correlations among the baths and the optimal performance of the engine.

It is worth noticing that, despite the fact that the mutual information between different elements of the system can be substantially large, for our choice of parameters, none of the correlations built involve entanglement. Indeed, it is well known that entanglement in quantum fields decays very rapidly with temperature, reaching zero at a finite value \cite{Brown_2013}. However, for a sufficiently cold bath, one could still expect some entanglement to be present, although it is not clear whether it will play a significant role in the engine's performance.

\section{Summary and conclusions}
\label{sec:conclusions}

Using the formalism of Gaussian quantum mechanics we have been able to circumvent the standard assumptions of weak coupling, slow driving, and infinite size of the baths, usually employed in studying thermodynamic phenomena. The focus of our study was on a single, driven harmonic oscillator undergoing an Otto cycle between two finite harmonic thermal reservoirs. Despite the attention the physics beyond these assumptions has received in recent years, to the best of the authors' knowledge, this is the first work where none of these assumptions is made.

We first study the interaction of a machine with a single bath, modeled as an $N$-mode translationally-invariant harmonic ring. GQM allows to observe not only how the machine thermalizes, but also how the interaction creates correlations between the WM and the region of the bath that directly interacts with it, and how these correlations later on propagate across the bath. 

In our study of the quantum Otto cycle, we conclude that the crucial element that determines the performance of the engine is the propagation of the perturbations created by the WM-bath interaction. During the first cycles of operation, the WM interacts with the baths in such a way that an infinite-sized-baths behaviour is observed. We call these cycles ``perfect'' and find that the figures of merit of the engine during these cycles remain almost unchanged with the efficiency being very close to its optimal value, while maintaining finite power and respecting the power-efficiency trade-off. We furthermore observe that the perturbations generated during these interactions propagate through the baths as wavepackets moving with constant velocity. These wavepackets leave residual perturbations behind, which causes a slow, but gradually-accumulating degradation of the engine's performance ---an effect that persists even for infinitely large baths. After enough time, the wavepackets return to the interaction region and start disrupting the thermalization of the WM, thereby drastically affecting the work output and efficiency of the machine. We expect this picture to also hold beyond the Gaussian regime, provided the speed of sound within the baths is finite \cite{Nachtergaele_2009, Nachtergaele_2010}.

We have also explored the interplay between the degradation of our engine and the creation of correlations within the overall system. As discussed, the process of running the engine inevitably creates an increasing number of correlations with the baths, within the baths, and among the baths. Remarkably, the bath-bath correlations remain very close to zero during the perfect cycles, and start increasing abruptly right after. This represents the overall system's gradual evolution to a more and more passive state. We believe that further study into this interesting dynamics is warranted.

By its own example, this work demonstrates the capabilities of Gaussian quantum mechanics as a workhorse for assessing finiteness effects in a field that has historically relied on infinite (time, size, and subtlety of the interactions) idealizations. Not only can GQM address fundamental questions in quantum thermodynamics, but it also provides us with more tractable numerical computations. This we believe may be of great use for the community of quantum thermodynamics.

\textit{Note added} -- After having submitted this manuscript, we became aware of related work \cite{Reid_2017}, which studies a finite-size Gaussian engine consisting of a single-oscillator working medium operating an Otto cycle between two heat baths composed of single oscillators each.

\acknowledgments

We thank Ronnie Kosloff for useful comments. \mbox{A. P.-K.} gratefully acknowledges Fundaci\'on Obra \mbox{Social} \mbox{``la Caixa''} for their support. The work of \mbox{K. V. H.} was supported by the Villum Fonden. \mbox{E. G. B.} acknowledges the support of the Natural Sciences and Engineering Research Council of Canada. All authors acknowledge financial support from the Spanish MINECO (QIBEQI \mbox{FIS2016-80773-P} and Severo Ochoa SEV-2015-0522), Fundaci\'o Privada Cellex and the Generalitat de Catalunya (SGR875 and CERCA Program).

\bibliographystyle{apsk}

\bibliography{references}

\appendix

\clearpage

\section{The effective temperature of the WM during an isochoric interaction} \label{app:temperature}

Due to the strong coupling to the bath during the isochoric interaction, the state of the WM will in general acquire non-diagonal terms, leading to the state being, in general, not a thermal state. Indeed, the thermal state is a function of the Hamiltonian and hence cannot have non-diagonal terms in the energy eigenbasis. In this appendix, we introduce a measure of athermality for Gaussian states of an oscillator and show that for the isochoric interactions we consider in the main text (Secs.~\ref{sec:interaction_with_single_bath} and \ref{sec:Otto}) the athermality is negligible, especially at the end of the interaction. This additionally justifies the view that the thermal bath thermalizes the WM. 

Let a single-oscillator Gaussian state be described by a covariance matrix $\bs \sigma_m=\left(\begin{array}{cc} \nu_1 & \kappa \\ \kappa & \nu_2 \end{array}\right)$. If the state were thermal, then its covariance matrix would simply be $\bs{\hat{\sigma}}_m=\left(\begin{array}{cc} \nu & 0 \\ 0 & \nu \end{array}\right)$, i.e., $\nu_1=\nu_2\equiv\nu$ and $\kappa=0$. Then, following Eq.~\eqref{GroundThermal}, its temperature would be
\bea
T = \frac{\omega_m}{\ln\frac{\nu+1}{\nu-1}}.
\label{Teff}
\eea

This will not be, however, the case of our WM for every instant during the interaction with a bath. In order to prescribe a temperature to a general state of the WM we take its covariance matrix, $\bs \sigma_m$, symplectically diagonalize it to $\bs{\hat{\sigma}}_m=\left(\begin{array}{cc} \tilde{\nu} & 0 \\ 0 & \tilde{\nu} \end{array}\right)$ as prescribed in Sec.~\ref{sec:GQM}, and compute its temperature via Eq.~\eqref{Teff}. The obtained temperature will be the effective temperature of the original state described by $\bs \sigma_m$. This is the definition we use in Fig.~\ref{fig:transfer}. Note that, for thermal states, this effective temperature coincides with the real temperature of the system.

In order to define the athermality of the state given by $\bs\sigma_m$, $\rho(\bs \sigma_m)$, we compute the Uhlmann fidelity \cite{mikeike} between $\rho(\bs\sigma_m)$ and the thermal state at the effective temperature, $\rho(\bs{\hat{\sigma}}_m)$, given by
\begin{equation}
F[\bs\sigma_m,\bs{\hat{\sigma}}_m]=\left( \tr\sqrt{\sqrt{\rho(\bs\sigma_m)} \rho(\bs{\hat{\sigma}}_m) \sqrt{\rho(\bs\sigma_m)}} \right)^2.
\end{equation}

$F[\bs\sigma_m,\bs{\hat{\sigma}}_m]$ is equal to $1$ if and only if $\bs\sigma_m=\bs{\hat{\sigma}}_m$, and is $<1$ otherwise.

For Gaussian states the Uhlmann fidelity can be directly expressed in terms of their covariance matrices. For purely quadratic states (which recall is the case of thermal states) the fidelity is given by \cite{Scutaru_1998}
\bea \label{scut}
F[\bs\sigma_m,\bs{\hat{\sigma}}_m]=\frac{2}{\sqrt{A+B}-\sqrt{B}},
\eea
where the quantities $A$ and $B$ are given by
\begin{align}
    A&=4\det(\bs\sigma_m+\bs{\hat{\sigma}}_m),\\ B&=(4\det\bs\sigma_m-1)(4\det\bs{\hat{\sigma}}_m-1).
\end{align} 

Now we can look at how much the state of the WM, as given by the covariance matrix $\bs\sigma_m(t)$, differs from the thermal state at temperature $T_{eff}$---given by the covariance matrix $\bs{\hat{\sigma}}_m(t)$---at any moment $t$ during the isochoric interactions described in Secs.~\ref{sec:interaction_with_single_bath} and \ref{sec:Otto}. To that end, we define the athermality of the state at a moment $t$ as
\bea
\mathcal{A}(t)=1-F[\bs\sigma_m(t),\bs{\hat{\sigma}}_m(t)].
\eea

Due to the properties of the fidelity, we thus have that $0\leq\mathcal{A}(t)\leq 1$ and $\mathcal{A}(t)=0$ iff $\bs\sigma_m(t)=\bs{\hat{\sigma}}_m(t)$, i.e., if the state of the system is really thermal.

\begin{figure}
    \centering
    \includegraphics[width=0.48\textwidth]{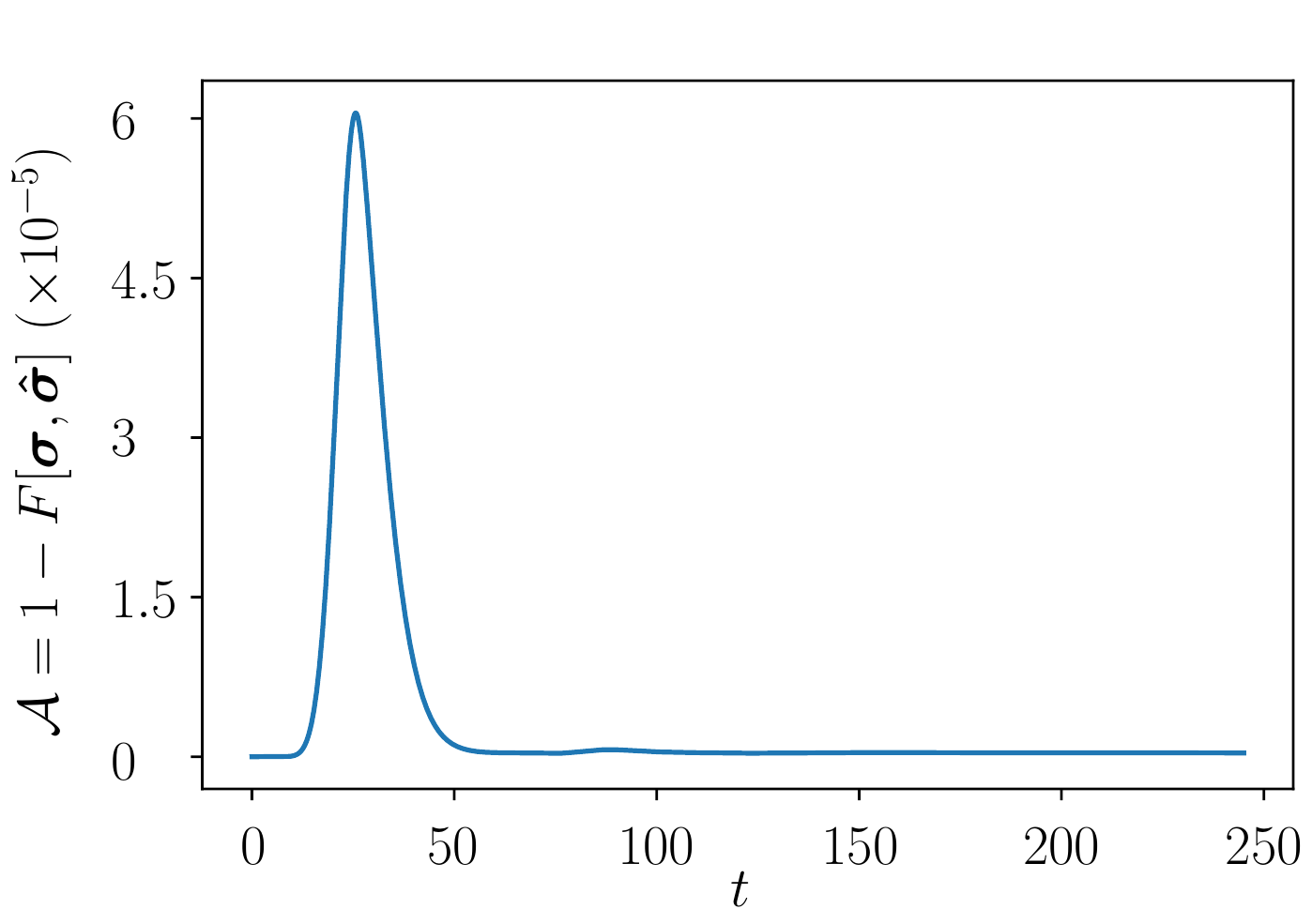}
    \caption{Time-dependence of the athermality of the WM, initially thermal at temperature $T_b=0.5$, for the isochoric interaction with a bath (at temperature $T_b=4$) consisting of $N=30$ oscillators with frequencies $\omega_b=\omega_m=2$. The coupling constants are matched so that $\gamma=\alpha=0.1$ and the interaction lasts $\tau=245$ units of time, with the ramp-up time $\delta$ being $0.1\tau$. See Sec.~\ref{sec:interaction_with_single_bath} for a detailed explanation of the meaning of these parameters.}
    \label{fig:athermality}
\end{figure}

In Fig.~\ref{fig:athermality} we plot the athermality of the WM for an example isochoric interaction with a bath consisting of $N=30$ oscillators. We clearly see that, except for a period in the beginning of the process, the WM is rather close to being thermal. It is especially interesting that by the end of the process, the WM is almost exactly thermal. This information, in addition to the final temperature of the WM shown in Fig.~\ref{fig:transfer}, indicates that the WM is thermalized by the bath.

\section{Adiabat without mass change} \label{app:swap}

Instead of performing the Hamiltonian swap in Eq.~\eqref{swap}, which is equivalent to simultaneously quenching both the frequency and the mass of the WM, let us explore the possibility of changing only the frequency of the WM. Namely, 
\bea \label{adi1}
H_m=\frac{P_m^2}{2\mu}\!+\!\frac{\mu\omega_m^2Q_m^2}{2} \leftrightarrow H'_m=\frac{P_m^2}{2\mu}\!+\!\frac{\mu(\omega'_m)^2Q_m^2}{2},~~~~
\eea
where $Q_m$ and $P_m$ are the canonical position and momentum of the WM, and $\mu$ is its mass. In this case, the quadratures and the creation-annihilation operators change as well:
\bea \label{adi2}
q_m'=Q_m\sqrt{\mu\omega_m'},\,\,\,
a_m'=\sqrt{\frac{\mu\omega_m'}{2}}\!\left(\!Q_m\!+\! \frac{\mathrm{i}}{\mu\omega_m'} P_m\!\right)\!.~~~
\eea

In these terms, the change in Eq.~\eqref{adi1} takes the form
\bea \label{adi3}
H_m=\omega_ma_m^\dagger a_m \leftrightarrow H'_m=\omega_m' (a'_m)^\dagger a'_m.
\eea

Now, if we perform this change instantaneously, the initial thermal state of the WM, $\rho\propto e^{-H_m / T_m}$, will remain unchanged. Since $[H_m',H_m]\neq 0$, this will result in the state having coherences in the new energy eigenbasis \cite{Kosloff_2017}, thereby significantly reducing the efficiency of the engine (this can be straightforwardly deduced from the analysis presented in Appendix~\ref{app:eff}). If, on the other hand, as a result of Eq.~\eqref{adi1}, the state were also changed to
\bea \label{adi4}
\rho'\propto \exp\left(-\omega_m (a'_m)^\dagger a'_m / T_m\right),
\eea
its covariance matrix, as defined by Eq.~\eqref{covmat} with $\bs x_m'$ instead of $\bs x_m$, would remain unchanged. If now the interaction Hamiltonian, $H_{int}$ (see Eq.~\eqref{Hint}), would couple to the bath degrees of freedom with the new quadrature, $q_m'$ (instead of $q_m$), the dynamics of the overall covariance matrix would be the same as that presented in the main text. 

Put in other words, were we to change to the new quadratures in all formulas, the dynamics on the level of covariance matrices and Hamiltonian matrices would remain unchanged. Hence, we would have exactly the same results for such an engine that those shown in the main text.

For such a modification to work, we need to have $\rho$ evolving into $\rho'$ as a result of Eq.~\eqref{adi3}. Since the instantaneous change in Hamiltonian cannot produce any changes in the state, let us allow the change to take some non-zero time $\tau_{ad}$. The problem is now the following: is it possible to devise a quadratic Hamiltonian path connecting $H_m$ to $H'_m$ that is capable of evolving the state in time $\tau_{ad}$ so that the covariance matrix remains the same?

To answer the above question, let us notice that the covariance matrix remains unchanged as a result of quantum adiabatic evolution. The latter is defined by the unitary evolution operator
\bea \label{adi5}
U=\sum_n |n\rangle_{{}_{H_{m}'}} \langle n|_{{}_{H_{m}}},
\eea
where $|n\rangle_{{}_{H_{m}}}$ and $|n\rangle_{{}_{H_{m}'}}$ are the $n$-th eigenvalues of, respectively, $H_m$ and $H_m'$ (see, for instance, Ref. \cite{Muga_2010}). Indeed, we have that
\begin{align}
\sigma'_{ab}=&\tr(U\rho U^\dagger\,(x'_a x'_b+x'_bx'_a))\notag\\
=&\tr(\rho\,(x_a x_b+x_bx_a)),
\end{align}
where we used the fact that
\begin{align}
    U^\dagger \bs x' U=&\sum_{n,k}|n\rangle_{{}_{H_{m}}}\langle k|_{{}_{H_{m}}}\cdot \langle n|_{{}_{H_{m}'}}\bs x'|k\rangle_{{}_{H_{m}'}}\notag\\
    =&\sum_{n,k}|n\rangle_{{}_{H_{m}}}\langle k|_{{}_{H_{m}}} \cdot \langle n|_{{}_{H_{m}}}\bs x|k\rangle_{{}_{H_{m}}}\notag\\
    =&\,\bs x.
\end{align}

Now, as is shown in Ref.~\cite{Muga_2010}, for a single oscillator, a shortcut to adiabaticity can be constructed for implementing the unitary in \eqref{adi5} by adding the time-dependent term
\bea \label{adi7}
H_I(t)=-\frac{\dot{\omega}_m(t)}{4\omega_m(t)}(QP+PQ)
\eea
to the Hamiltonian of the oscillator for the period of the adiabat $\tau_{ad}$. Here, the function $\omega_m(t)$ is arbitrary provided that it satisfies $\omega_m(0)=\omega_m$ and $\omega_m(\tau_{ad})=\omega'_m$. 

In order for this new cycle to coincide with the one in the main text, we need $\tau_{ad}$ to approach to zero. This would require very fast generation of the term \eqref{adi7} and very quick driving of the frequency. Although we leave the question of the experimental accessibility of such quick driving open, we note that the codes provided in the computational appendix \cite{compapp} straightforwardly allow for simulating an Otto cycle with any non-zero $\tau_{ad}$.

\section{Full-cycle energetics} \label{app:eff}

In this appendix, we perform a detailed analysis of the work and heat involved in the perfect cycles of operation of the WM. As in the main text, we take as a starting point the moment of the cycle where the WM is still in a state $\propto e^{-\omega_c a^\dagger_m a_m/T^{(0)}_c}$ but its Hamiltonian has already been changed to $\omega_h a^\dagger_m a_m$, and the isochoric interaction with the hot bath is about to start. We label the initial temperature of the bath with superscript $^{(0)}$ to indicate that it was its temperature before the first cycle started. At this moment, the total Hamiltonian is just the sum of the individual Hamiltonians, and the energy before starting the isochore is thus
\bea
E^{(in)}=E_B^{(in)}+E_{W\!M}^{(in)}=E_B^{(in)}+\omega_h\n{\frac{\omega_c}{T^{(0)}_c}},
\eea
where we have defined
\bea
n(x)=\frac{1}{e^x-1}.
\eea

After the isochore, the total Hamiltonian returns to being the sum of the individual Hamiltonians and the state of the WM is again thermal (albeit now correlated with the bath). Hence, the energy of the system right after the isochore is
\bea
E^{(fin)}=E_B^{(fin)}+\omega_h \n{\frac{\omega_h}{T^{(1)}_h}},
\eea
where $T^{(1)}_h$ is the temperature of the WM after the interaction with the hot bath. Recall that, as discussed in Section~\ref{sec:Otto:performance}, although the parameters can be chosen so that $T_h^{(1)}=T_h$ exactly, this temperature does not need to be exactly the temperature of the bath $T_h$. In the example in the main text, namely, when $N_c=N_h=30$, $\omega_c=1$, $\omega_h=2$, $T_c=0.5$, $T_h=4$ and $\tau=100$, $T_h^{(1)}$ is slightly greater than $T_h$: $T_h^{(1)}-T_h\approx 1.3\times 10^{-4}$. This small difference does not affect our analysis because such deviations, if not appearing in the first cycle, do appear in the subsequent ones. The work extracted during the hot isochore is then
\begin{equation}
    W_{ih} \!=\! E^{(in)} \!-\! E^{(fin)} \!=\! Q + \omega_h \hspace{-0.5mm} \left[ \n{\frac{\omega_c}{T^{(0)}_c}} \!-\! \n{\frac{\omega_h}{T^{(1)}_h}} \! \right],
\end{equation}
where $Q=E^{(in)}_B - E^{(fin)}_B$.

Rearranging and adding the superscript $^{(1)}$ to indicate that the labeled quantities correspond to the end of the first cycle, we obtain the following expression for the heat exchanged during the cycle:
\bea
Q^{(1)} = \omega_h \left[\n{\frac{\omega_h}{T^{(1)}_h}} \!-\! \n{\frac{\omega_c}{T^{(0)}_c}} \! \right] + W^{(1)}_{ih}.~~
\eea

The following step in the cycle is the adiabatic expansion. The work extracted from this process is
\bea
W^{(1)}_{h\to c}=(\omega_h-\omega_c)\n{\frac{\omega_h}{T^{(1)}_h}}.
\eea

Next, during the cold isochoric interaction, we extract $W^{(1)}_{ic}$ amount of work, and leave the system at temperature $T^{(1)}_c$ (which, again, is slightly different from $T^{(0)}_c=T_c$. For the case studied in section \ref{sec:Otto}, \mbox{$T^{(1)}_c-T^{(0)}_c\approx 7.7\times 10^{-3}$}).

Finally, during the adiabatic compression, we extract the negative amount of work
\bea
W^{(1)}_{c\to h}=-(\omega_h-\omega_c)\n{\frac{\omega_c}{T^{(1)}_c}}.
\eea

The final state of the WM will be $\propto e^{-\omega_c a^\dagger_m a_m/T_c^{(1)}}$, which explicitly shows that the WM is not completely cyclic. However, the deviations from cyclicity, in the case discussed in the main text, are of \mbox{$\mathcal{O}(e^{-\omega_c/T_c^{(0)}}-e^{-\omega_c/T_c^{(1)}})=\mathcal{O}(10^{-3})$}, the same order of magnitude of the degradation during the perfect cycles.

The total work output of the cycle is the sum of the outputs in every step, that is
\begin{equation}
W^{(1)}\!=\!(\omega_h \!-\! \omega_c) \! \left[\n{\frac{\omega_h}{T^{(1)}_h}}-n\Bigg(\frac{\omega_c}{T^{(1)}_c}\Bigg) \right] \!+\! W^{(1)}_{ih} \!+\! W^{(1)}_{ic},
\end{equation}
and thus the efficiency, $\eta^{(1)}=W^{(1)}/Q^{(1)}$, will amount to
\begin{align}
    \eta^{(1)} = \eta_O &+ \frac{\omega_c W^{(1)}_{ih}+\omega_h W^{(1)}_{ic}}{\omega_h Q^{(1)}} \notag\\
    &-\frac{\omega_h-\omega_c}{Q^{(1)}}\left[ \n{\frac{\omega_c}{T^{(1)}_c}}-\n{\frac{\omega_c}{T^{(0)}_c}} \right],
    \label{appeff2}
\end{align}
where
\bea
\eta_O=1-\frac{\omega_c}{\omega_h}
\eea
is the maximal theoretical efficiency for the Otto engine in which the WM couples negligibly weakly to infinite, Markovian baths \cite{Rezek_2006, Kosloff_2017}. Since $T_c^{(1)}-T_c^{(0)}\ll 1$, whenever $\eta_O$ is away from the Carnot value, namely, when
\bea
\frac{\omega_h}{T^{(1)}_h}<\frac{\omega_c}{T^{(0)}_c}
\eea
so that
\bea
Q^{(1)}\gg \max\left\{ W^{(1)}_{ih}, W^{(1)}_{ic} \right\},
\eea
the efficiency $\eta^{(1)}$ will be very close to $\eta_O$. In our example, both the hot and cold isochoric works are of the order of $10^{-3}$, while $Q^{(1)}\approx 3$.

By the moment the second hot isochore (and hence the second cycle) is about to start, the perturbations created in the hot bath during the first isochore will have traveled away from the interaction node in form of a wavepacket. However, the propagation of this wavepacket is not ideal in that it leaves a trace in the form of residual perturbations. In particular, before the start of the second isochore, the state of the interaction node will be slightly different from that at equilibrium. This means that $T_h^{(2)}$ will be even further from $T_h$ than $T_h^{(1)}$. As our numerical analysis shows for perfect cycles, and as it is to be expected from the fact that the hot isochore extracts heat from the bath,
\bea \label{degrad1}
T_h \equiv T_h^{(0)}\approx T_h^{(1)}\gtrsim T_h^{(2)}\gtrsim \cdots.
\eea

With a similar reasoning, another result that we observe numerically is that
\bea \label{degrad2}
T_c \equiv T_c^{(0)} \approx T_c^{(1)}\lesssim T_c^{(2)}\lesssim \cdots.
\eea

Moreover, since (while within perfect cycles) the WM is thermal after each interaction with the baths, for the $k$-th perfect cycle we have that
\bea \label{qk}
Q^{(k)} = \omega_h \left[\n{\frac{\omega_h}{T^{(k)}_h}} \!-\! \n{\frac{\omega_c}{T^{(k-1)}_c}} \! \right] + W^{(k)}_{ih}~~
\eea
and
\bea \nonumber
W^{(k)} \!=\! (\omega_h \!-\! \omega_c) \! \left[\n{\frac{\omega_h}{T^{(k)}_h}}-\n{\frac{\omega_c}{T^{(k)}_c}} \right] \!+\! W^{(k)}_{ih} \!+\! W^{(k)}_{ic},
\\ \label{wk}
\eea
and therefore
\begin{align}
    \eta^{(k)} = \eta_O &+ \frac{\omega_c W^{(k)}_{ih}+\omega_h W^{(k)}_{ic}}{\omega_h Q^{(k)}} \notag\\
    &-\frac{\omega_h-\omega_c}{Q^{(k)}}\left[ \n{\frac{\omega_c}{T^{(k)}_c}}-\n{\frac{\omega_c}{T^{(k-1)}_c}} \right].
    \label{etak}
\end{align}

Along with Eqs.~\eqref{degrad1} and \eqref{degrad2}, Eqs.~\eqref{qk} and \eqref{wk} explain the slow, gradual decrease of cycle heat and work during the period of perfect operation (see Fig.~\ref{fig:wadiabatic}). At the same time, Eq.~\eqref{etak} explains why is that the efficiency does not accumulate errors and stays very close to $\eta_O$ throughout the perfect performance. Indeed, as mentioned above, the isochoric works, $W_{ih}^{(k)}$ and $W_{ic}^{(k)}$, stay of the order of $\alpha^2$ and the acyclicity, as given by $T_c^{(k)}-T_c^{(k-1)}$, being an effect a \emph{single} WM-bath interaction session has on the bath, remains almost unchanged throughout the perfect cycles and is small compared to the cycle heat. Another important consequence of Eq.~\eqref{etak} is that Eq.~\eqref{eficiencia} needs to be slightly modified for $k\geq 2$. Indeed, although $\eta^{(k)}_{\alpha=0}-\eta^{(k)}\propto \alpha^2$ still holds, one needs additionally account for
\begin{equation} \label{effcorr2}
\eta_O - \eta^{(k)}_{\alpha=0} = \mathcal{O}(T_c^{(k)}-T_c^{(k-1)}),
\end{equation}
and if, for the first cycle, this term can be eliminated by adjusting the interaction time, it will be non-zero for the subsequent cycles. However, as we mentioned above, the correction \eqref{effcorr2} is very small and does not increase as the cycles proceed.

Lastly, we remark that, whenever the machine approaches the Carnot efficiency, $\eta_C=1-T_c/T_h$, namely, when $\omega_c/\omega_h$ approaches $T_c/T_h$ from above, the work output of the engine tends to zero (as can be seen from Eq.~\eqref{wk}), and, exactly at the point when $\eta_O=\eta_C$, $W^{(k)}<0$.

\section{Relative entropy}\label{app:relentropy}

In order to show the evolution of the baths' states, we choose the relative entropy as a distance quantifier. Take, for example, the hot bath. Let us denote its state at the beginning of the $i$-th cycle by $\rho_h^{(i)}$. Then, the quantity we are interested in is
\bea \label{relent0}
S(\rho_h^{(i)} || \rho_h^{(1)}) = S\left( \rho_h^{(i)} \Big|\Big| \frac{1}{Z_h}e^{-H_h/T_h} \right),
\eea
where $H_h$ is the Hamiltonian of the hot bath, \mbox{$Z_h=\tr e^{-H_h/T_h}$}, and \cite{mikeike}
\begin{equation}
S(\rho||\sigma) = \tr\left[\rho(\ln\rho - \ln\sigma)\right].
\end{equation}

The relative entropy has several features desirable for a distance measure. In particular, those that are of interest for the case studied are that
\begin{equation}
S(\rho||\sigma)\geq 0,
\end{equation}

i.e., that it is a positive quantity, and that

\begin{equation}
S(\rho||\sigma)=0 \quad \text{iff} \quad \rho = \sigma.
\end{equation}

Although the relative entropy does not satisfy the triangle inequality and is not symmetric, which means it is not a distance measure in the proper sense, it decreases monotonically under completely positive trace preserving operations \cite{mikeike}, which makes it a distinguishability measure of choice in many situations \cite{mikeike}.

Whenever the second argument in $S(\bullet||\bullet)$ is a Gibbs state, as, e.g., is the case in Eq.~\eqref{relent0},
\bea \label{relent1}
S(\rho_h^{(i)} || \rho_h^{(1)}) \! &=& \! T_h^{-1}\!\left(E_h^{(i)} \!-\! E_h^{(1)}\right) \!-\! S(\rho_h^{(i)}) \!+\! S(\rho_h^{(1)}) ~~~~~
\\ \label{relent2}
&=& \! T_h^{-1} F[\rho_h^{(i)}] - T_h^{-1} F[\rho_h^{(1)}],~~~~
\eea
where $S$ is the von Neumann entropy, $E_h^{(i)}$ is the energy of the hot bath at the beginning of the $i$-th cycle, and $F = E-TS$ is the free energy. Eq.~\eqref{relent2} means that relative entropy measures the ``thermodynamic'' distance between $\rho_h^{(i)}$ and $\rho_h^{(1)}$, which additionally motivates our choice of relative entropy as a distance quantifier.

The quantities in Eq.~\eqref{relent1} can be readily calculated directly in the covariance-matrix picture via Eqs.~\eqref{energyEqn}, \eqref{vnent0}, and \eqref{vnent1}. An example of such a calculation is presented in Fig.~\ref{fig:wadiabatic}b, where it can be seen that the relative entropy distance of the state of of the bath from the initial state of the bath increases at a steady rate as the ``perfect'' cycles proceed. This is contrasted by how the efficiency and work per cycle remain almost constant during these cycles.

\end{document}